\documentclass[openacc]{rsproca_new}
\usepackage{comment}

\titlehead{Research}

\begin{document}

\title{Linear Response and Optimal Fingerprinting for Nonautonomous Systems}

\author{
Valerio Lucarini$^{1,2}$}

\address{$^{1}$ School of Computing and Mathematical Sciences, University of Leicester, Leicester, UK\\
$^{2}$School of Sciences, Great Bay University, Dongguan, P.R. China}

\subject{Applied Mathematics, Statistical Physics, Mathematical Modeling}


\corres{Valerio Lucarini\\
\email{v.lucarini@leicester.ac.uk}}

\begin{abstract}
We provide a link between response theory, pullback measures, and optimal fingerprinting method that paves the way for a) predicting the impact of acting forcings on time-dependent systems and b) attributing observed anomalies to acting forcings when the reference state is not time-independent. We  derive formulas for linear response theory for time-dependent Markov chains and diffusion processes. We discuss existence, uniqueness, and differentiability of the equivariant measure under general (not necessarily slow or periodic) perturbations of the transition kernels. Our results allow for extending the theory of optimal fingerprinting for detection and attribution of climate change (or change in any complex system) when the background state is time-dependent and when the optimal solution is sought for multiple time slices at the same time. We provide  numerical support for the findings by applying our theory to a modified version of the Ghil-Sellers energy balance model. We verify the precision of response theory - even in a coarse-grained setting - in predicting the impact of increasing CO$_2$ concentration on the temperature field. Additionally, we show that the optimal fingerprinting method developed here is capable to attribute the climate change signal to multiple acting forcings across a vast time horizon.
\end{abstract}
\maketitle


\section{Introduction}
Response theory describes how the statistics of the system of interest are impacted by the application of a (weak) external forcing, which can in general depend on time. The overall goal is to provide formulas whereby the response operators can be written in terms of the statistical properties of the system in the unperturbed state and of the nature of the forcing. Additionally, response theory aims at better characterising the notion of \textit{weak forcing}, i.e., by identifying under which conditions the response operators diverge. Under specific conditions associated with the emergence of criticality, even infinitesimal perturbations can lead to tipping behaviour.  

Specifically, the fluctuation-dissipation theorem (FDT) \cite{Kubo1957,Kubo1966} establishes that, when taking the approximation of a linear relation between the intensity of the forcing and the amplitude of the response, the response operators describing the time-dependent departure of the ensemble averages of observables with respect to the reference statistics can be written as correlations of suitably defined observables in the unperturbed state. 

Whilst the FDT was originally established in the fairly restrictive scenario of thermodynamic systems that are perturbed from their reference canonical ensemble, it has been shown that the theorem can be substantially generalised \cite{HairerMajda2010}, including the case of systems whose reference state is far from equilibrium \cite{Lippiello2005,Marconi2008,Baiesi2009,Baiesi2013,Colangeli_2014,RespTheoryVulp,Wormell2019}. Additionally, response formulas can be extended to deal with nonlinear terms as well as for studying the change in higher order statistical moments. A hierarchical structure is apparent: the higher the order of nonlinearity and/or the statistical moment one is interested in, the more convoluted (in terms of involving higher order correlations) the response operators \cite{ruelle_nonequilibrium_1998,Bouchaud2005,Lippiello2008,Lucarini2008,LC12,Basu2015,lucariniwouters}.

Note that whilst it is possible to develop a response theory for special classes of deterministic chaotic systems possessing an invariant measure that is singular with respect to Lebesgue, the response operators cannot be cast in the form of correlation functions evaluated on the unperturbed state, so that the standard form of FDT does not apply \cite{Ruelle1998GeneralLinearresponseformula}. The lack of a direct link between forced and free fluctuations is due to the fact that as a result of the contraction of the phase space the fluctuations of the system do not involve excursions along the stable directions in the tangent space, whilst the forced motions have in general a non-vanishing projection on the stable, the unstable, and the neutral directions \cite{Ruelle2009}. As a result of the very different statistical and dynamical properties of the system in the stable vs unstable directions of the tangent space it has proven extremely challenging to develop algorithms for computing the response operators for chaotic systems  \cite{abramov2007,Ni2020,Chandramoorthy2022,Ni2023}.

A step forward in our ability of linking forced and free fluctuations of a general system comes from combining response theory with Koopmanism \cite{Mezic2005,Budisic2012,Brunton2022}. By suitably inserting projectors in the subspaces spanned by the various Kolmogorov modes of the unperturbed system in the response operators, one gains interpretability, because the linear response operator can be expressed as a sum of terms, each associated with a specific mode of variability of the reference system \cite{Santos2022,LucariniChekroun2023,lucarini2025generalframeworklinkingfree,zagli_SIAM:2026}. The results extend to the nonlinear case where the nonlinear response operators are broken into terms that describe the interaction of the forcing with multiple modes of variability of the system. Such a spectral approach delivers formulas that are visually identical, e.g. to those associated with perturbation theory for quantum mechanical systems, and makes it possible to extend response theory for the case of mixed jump-diffusion processes \cite{Chekroun_zagli_lucarini_2025}.

The spectral approach also clarifies that critical behaviour emerges when the decay of correlation between generic observables becomes subexponential \cite{chekroun2019c}, as a result of the closure of the so-called spectral gap, which is associated with the real part of the first subdominant eigenvalue of the Kolmogorov operator. This provides a solid foundation \cite{LucariniChekroun2024} to the theory of  critical slowing down, which was first introduced in the context of second order
phase transitions \cite{Hohenberg1977} and then widely used in the context of tipping points research \cite{scheffer2009early,lenton2012,dakos2024tipping}.

The practical implementation of the spectral response theory  relies on the choice of the Kolmogorov dictionary \cite{lucarini2025generalframeworklinkingfree}, and currently benefits from the extremely encouraging  development of accurate, efficient, and mathematically sound variants of the so-called extended dynamical model decomposition (eDMD) \cite{Colbrook2023_MultiverseDMD,Colbrook_ResDMD_Stochastics,Colbrook_ResDMD}. A given Kolmogorov dictionary provides a finite basis for optimally estimating the Kolmogorov modes of the system. Whilst explicit dictionaries like tensor products of monomials or of Chebyshev polynomials suffer from the curse of dimensionality, kernelised version of eDMD (keDMD) \cite{Williams2016,Klus2020} has been shown to perform very well also for high-dimensional systems. 

A special case of Kolmogorov dictionary is given by characteristic functions defined on a partition of the phase space, as, e.g., those resulting from  a Voronoi tessellation \cite{Aurenhammer91} associated with $N$ centers. By supplementing such a spatial coarse-graining with a suitable temporal coarse- graining whereby the time is discretised, the evolution of the system is represented as a finite-state Markov chain \cite{Nor97}, where the stochastic matrix can be estimated from data only, without the need to know the exact form of the evolution equation of the system, Optimising this coarse-graining procedure is central to Markov state modelling \cite{Pande2010,Schuette2012,Husic2019}, where the initial partition is usually constructed by applying k-means clustering \cite{Macqueen1967} to the complete dataset. This procedure is  equation-agnostic and also able to deal with high-dimensional systems. This coarse-grained angle on the dynamics is particularly fruitful also when considering the response of the system to perturbations, because response theory for Markov chains is crystal clear in terms of the domain of applicability - one can clearly define the radius of convergence of perturbative expansions - and in terms of explicit formulas for linear and nonlinear response operators, which can be expressed in relatively simple matrix relations \cite{Baiesi2009JSP,Lucarini2016,Santos2020,Aslyamov2024,Lucarini2025,Aslyamov2025}. The practical use of such formulas is greatly favoured by the availability of extremely efficient and accurate scientific software that can deal with very large matrices even on relatively mundane computer facilities \cite{MATLAB2024,PYTHON,JULIA}.

A separate route for achieving via data-driven methods accurate estimates of the linear response operators has been recently proposed by Giorgini and collaborators \cite{giorgini2024linear, giorgini2024datadriven,giorgini2025statistical}, who combined generative modelling with the FDT  and managed to estimate, in some relevant examples, how chaotic dynamical systems and stochastic models respond to perturbations.

Linear response is targeted at making statements on ensemble averages. Yet, it can be helpful also for studying key properties of individual trajectories of a system. In a previous contribution we have shown how response theory provides the dynamical foundation for the optimal fingerprinting method (OFM) for detection and attribution of climate change \cite{LucariniChekroun2024}. 

The OFM is a statistical method for (optimally, in a least squares sense) associating the observed climate change signal to different fingerprints, each related to a specific forcing impacting the climate system. OFM has been essential for establishing key results in climate change science and particularly in allowing us to attribute the observed ongoing climate change (mainly) to anthropogenic causes as well as to recognise the impact of natural forcings \cite{Hasselmann1997,Allen1999,Hegerl2011,Hannart2014}. One of the cornerstones of OFM, namely Pearl's notion of causality based on the interventionist approach \cite{Pearl2009}, which is the base of the definition of fingerprints  finds concrete expression in causality of the Green's functions, which have been shown in \cite{LucariniChekroun2024} to be the building blocks of the fingerprints. Our previous results have also clarified how to extend the OFM scope and applicability to nonlinear effects, and clarified that the OFM tests the observed climate change signal against the null hypothesis of a climate at steady state \cite{LucariniChekroun2024}. 

\subsection{Non-autonomous Dynamics}
The vast majority of results pertaining to response theory are based on the assumption that the reference state obeys autonomous dynamics, and that such a reference state is statistically described by the associated invariant measure. The response of the system is evaluated in terms of time-dependent changes of the measure of the forced system (or, alternatively, of one or more observables of interest) with respect to such statistical equilibrium.  Notable contributions in the direction of extending fluctuation-dissipation results and linear response to aging systems have been presented in the literature, see e.g. \cite{Crisanti2003,Bertier2007}.

Nonetheless, many systems of interest have a reference state that undergoes a periodic forcing or even more general aperiodic forcing, where the time modulation is such that no time-scale separation assumption can be made, as opposed to the case of aging media.  Examples of such systems include temporal networks \cite{Holmes2012,Vespignani2012,Ginestra2023}, such as social \cite{Ginestra2010} and spatial networks \cite{Ginestra2024}, time-dependent consensus models \cite{Leonie2021}, financial markets with time-varying volatility \cite{Chakrabarti2021}, ecosystems that are subjected to periodic forcing \cite{Summers2000,Jansen2000,Basak2024}, neural populations with time-dependent input \cite{Horrocks2024,Bolelli2025}, and non-stationary models of the economy \cite{Toscani2006,Kohlrausch2024}. 

In general, the problem arises when we want to study multiscale systems with no time-scale separation  and are able to describe theoretically or through numerical modelling only a limited range of scales. Explicit time dependence will unavoidably emerge from the effect of the unresolved time scales. This issue is particularly acute in the case of climate science \cite{Peixoto1992,saltzman_dynamical,GhilLucarini2020,LucariniChekroun2023} where variability resulting from a wealth of astronomical, astrophysical, and other natural factors exists on many orders of magnitude of time scales \cite{Anna2021}. As a result, it is extremely hard to rigorously or even operationally define   a reference steady state, whilst, instead, one has to resort to extremely useful but indeed heuristic definitions such as \textit{preindustrial conditions} in order to study the ongoing anthropogenic climate change \cite{IPCC2021}. An accurate evaluation of the rate of ongoing climate change requires careful treatment of various components of  climate variability \cite{Foster2011}, with the matter becoming particularly crucial as very recently evidence is emerging that climate change is, ominously, accelerating \cite{Foster2026}. 

The entire building of the previously mentioned OFM \cite{Hasselmann1997,Allen1999,Hegerl2011,Hannart2014} is based on the definition of a reference baseline steady state, on top of which the impact of natural and anthropogenic factors is observed. One might instead want to incorporate the natural forcings in a time-dependent baseline state, and target a more general OFM to anthropogenic forcings. Extending the OFM to the nonautonomous case would pave the way to causally connect observed anomaly signals with extra forcings acting on the system, thus allowing to better understand (and potentially control) ecosystems, financial markets, economic models, networks, and neural systems.  Note that the existence of explicit time dependence in the baseline dynamics requires also a rethinking of the notion of causality \cite{Sadeghi2025}.

\subsection{This Paper}
For time-dependent systems no stationary distribution exists in general. As discussed below, it is possible under suitable hypotheses, which boil down to a uniform contraction property, to introduce the notion of a unique equivariant measure, also known as a nonautonomous equilibrium \cite{Crauel1997,CHEKROUN20111685,GhilLucarini2020}, which leads to the existence of well-defined statistical properties for the system at all times. At a given time, one talks of \textit{snapshot} properties. In the case of nonautonomous deterministic dynamical systems, the notion of snapshot attractor \cite{Romeiras1990} has been widely exploited for conceptualising changing climate conditions from a dynamical systems perspective \cite{Bodai2013,Drotos2015,Herein2016,Tel2020,Bodai2020,Janosi2021}. 

 The purpose of this paper is to develop a complete linear response theory for time-dependent systems that is able to evaluate how adding additional (weak) forcing changes the properties of the system with respect to the time-changing reference state defined by the unperturbed dynamics. A key inspiration comes from the recent contribution by Branicki and Uda \cite{Branicki2021}, who were able to rigorously prove the existence of a response theory for suitably defined diffusion systems possessing time-periodic measures.  We aim here at deriving formal results that can be used for advancing our ability to investigate non-autonomous dynamical systems \cite{Ashwin2026,Crisan2026}. 

We  approach the problem using two separate yet related mathematical frameworks. We will derive explicit formulas for finite-state time-dependent Markov chains and for time-dependent diffusion processes. Our results do not rely on adiabatic or slow-variation assumptions. Instead, the response is expressed as a convergent infinite-time series that propagates the effect of perturbations forward along the dynamics. Special formulas are derived for the relevant case where the background dynamics undergoes a periodic modulation. Whilst there is a substantial conceptual overlap between the two mathematical frameworks and there is a clear correspondence between the obtained formulas, a key difference emerges in terms of applications, because Markov chains are especially suited for being used in a purely data-driven setting. 
 
We also show that the linear response theory developed here allows us to formulate the OFM also for time-dependent systems. Indeed, our new methodology allows us to associate signals of change of a system extracted from one individual trajectory to one or more extra acting forcings also in the case that the time-dependence of the reference system is very strong and can in principle be deployed for a large class of systems. Interestingly, the obtained OFM equations are almost identical to those used in the standard case where the reference system is at steady state. This further confirms the strength and robustness of the OFM.

Finally, we provide strong numerical support for the formal results presented in this paper by astudying a modified  version of the Ghil-Sellers energy balance model (EBM)\cite{Sellers,Ghil1976}, one of the foundational models of climate science. This consists of a reaction-diffusion one-dimensional partial differential equation that describes the evolution of the local energy budget of the climate system resulting from the competition of incoming solar radiation, outgoing infrared radiation, and horizontal energy transport associated with large-scale geophysical flows \cite{North1981,GhilLucarini2020}. We include a non-degenerate stochastic component to the evolution equation in order to take into account the effect of the unresolved degrees of freedom, along the lines of the Hasselmann program of stochastic climate models  \cite{Hasselmann1976,Imkeller2001,LucariniChekroun2023}.
We go conceptually beyond some of our earlier investigations \cite{LucariniChekroun2024} as we include explicit time dependence in the reference state as a result of considering - in a simplified way - the impact of the sunspot cycle and of volcanic forcings. We verify the accuracy of response theory in predicting the impact of increases of $CO_2$ in the temperature field even when we discretise the system using Markov state modelling approach.

Additionally, we consider a  more complex  scenario where we also include a simplified representation of the impact of increased aerosols in the mid-latitudes of the Northern Hemisphere which has a - mostly localised - cooling effect \cite{Recchia2023}. The aerosols forcing is mostly masked by the much stronger $CO_2$ forcing. Yet, using the OFM developed here one can attribute the climate change signal to the acting forcings thanks to their different spatial signatures.

The paper is structured as follows. In Sect. \ref{Markov} we derive response formulas for time-dependent Markov chains. In Sect. \ref{diffusion} corresponding results are obtained for diffusion processes. In Sect. \ref{optimalfingerprinting} we derive the optimal fingerprinting equations for time-dependent systems and discuss the solution of the problem in a simplified setting. In Sect. \ref{Ghil} we discuss the numerical simulations for the Ghil-Sellers model and use them to test response theory and optimal fingerprinting for time-dependent systems. In Sect. \ref{conclusions} we present our conclusions and provide perspectives for future work. Finally, Apps. \ref{periodicMarkov} and \ref{periodicdiffusion} provide specific results that apply in the case that the reference dynamics is periodic, and App. \ref{usualformulas} shows how to reconcile our key results with what is known from linear response for autonomous systems.

\section{Perturbations to time-dependent Markov Chains}\label{Markov}
Let $\mathcal{X} = \{1,\dots,N\}$ be a finite state space. We write $\mathcal{P}(\mathcal{X})$ for the probability simplex.
Let $\{\mathcal{M}_j\}_{j\in\mathbb{Z}}$ be a bi-infinite sequence of stochastic matrices, defining the time-dependent Markov chain process  as follows:
\begin{equation}
    \rho(j)=\mathcal{M}_{j-1}\rho(j-1), n\in\mathbb{Z}
\end{equation}
where $\rho(j)$ is the probability measure defined in $\mathcal{P}(\mathcal{X})$ obtained by evolving the the probability measure $\rho(j-1)$, which is also defined in $\mathcal{P}(\mathcal{X})$, using the stochastic matrix defined at time $j-1$. We assume that 
  $\forall j\in\mathbb{Z}$ and $\forall$ probability measures $\mu,\nu$ defined in $\mathcal{P}(\mathcal{X})$ the following applies:\[
\|\mathcal{M}_j\mu - \mathcal{M}_j\nu\|_{\mathrm{TV}} \le \delta \|\mu-\nu\|_{\mathrm{TV}}.
\]
where $0<\delta<1-\alpha$, $\alpha>0$ is the Dobrushin coefficient and $\|\mu\|_{\mathrm{TV}} = \frac12 \sum_{x\in\mathcal{X}} |\mu[x]|$ is the total variation norm. As a result, we have strong ergodicity and there is a unique sequence of probability measures $\nu(j)$ such that for any probability measure $\mu$ we have
\begin{equation}  \nu(j) = \lim_{m\rightarrow\infty} \mathcal{M}_{j-1}\mathcal{M}_{j-2}\ldots \mathcal{M}_{j-m} \mu. \label{covariantMarkov}
\end{equation}
The convergence to the limit sequence is exponential and defines the uniform contraction property of the sequence of matrices:
\begin{equation}
\|\mathcal{M}_j \mathcal{M}_{j-1} \ldots \mathcal{M}_1 \mu - \nu(n)\|_{\mathrm{TV}} \le C\delta^j \label{pullback}
\end{equation}
for any probability measure $\mu$. The sequence $\{\nu(j)\}_{j\in\mathbb{Z}}$ is the equivariant measure defining the statistical properties of the pullback attractor of the system, whilst $\nu(j)$ gives the measure of the snapshot attractor at time $j$. {Additionally, we have. that the expectation value of a measurable observable $\Phi:\mathcal{X}\to\mathbb{R}$  at time $j$ is given by
\begin{equation}
\langle \Phi \rangle(j)= \mathbb{E}_{\nu(j)}[\Phi]= \langle \Phi,\nu(j)\rangle = \Phi^\top\nu(j)= \sum_{x\in\mathcal{X}} \Phi[x] \nu(j)[x].  \label{observable0markov}
\end{equation}}
We have $\lim_{m\rightarrow\infty} \mathcal{M}_{j-1}\mathcal{M}_{j-2}\ldots \mathcal{M}_{j-m} \tilde{\mu}$ is the zero vector whenever $\tilde{\mu}$ belongs to the zero-mean subspace.  {The mathematical construction above generalises the spectral gap condition that is essential for developing response theory for autonomous Markov chains \cite{Lucarini2016,Santos2020,Lucarini2025}.}

Let $\mathcal{M}_j^\varepsilon = \mathcal{M}_j + \varepsilon m_j$, 
where each $m_j$ is a signed matrix such that for sufficiently small $\varepsilon$, $\forall j\in\mathbb{Z}$  $\mathcal{M}_j^\varepsilon$  remains stochastic and with Dobrushin coefficient bounded away from 1. {An especially relevant case is realised when $m_j=f(j)m$, \textit{i.e.} when the perturbation is given by time modulation of a fixed signed matrix $m$.} Let $\nu^\varepsilon(j)$ denote the associated equivariant measure.
As can be shown by a simple perturbation argument, the derivative $\nu^{(1)}(j)=\frac{d}{d\varepsilon}|_{\varepsilon=0} \nu^\varepsilon(j)$
exists and satisfies
\[
\nu^{(1)}(j) = \mathcal{M}_{j-1} \nu^{(1)}(j-1)+ m_{j-1} \nu(j-1)
\]
By using recursion and uniform contraction we can find an explicit expression for $\nu^{(1)}(j) $ as follows:
\begin{equation}
\nu^{(1)}(j) = \sum_{k=-\infty}^{\infty}\Theta(k) \mathcal{M}_{j-1}\ldots\mathcal{M}_{j-k} m_{j-k-1} \nu(j-k-1) \label{formulaMarkovmeasures}
\end{equation}
where $\Theta(k)=1$ if $k\geq0$ and $\Theta(k)=0$ otherwise. Note that the presence of $\Theta(k)$ ensures that the system obeys causality. 

We can translate the results above obtained for measures into something more practical by looking at the dual problem of studying how the expectation value of general measurable observables change as a result of the applied perturbation. Indeed, we obtain:
\begin{equation}
\frac{d}{d\varepsilon}\Big|_{\varepsilon=0} \mathbb{E}_{\nu^\varepsilon(j)}[\Phi] = \langle \Phi, \nu^{(1)}(j)\rangle = \sum_{k=-\infty}^{\infty}\Theta(k) \langle m_{j-k-1}^\top \mathcal{M}_{j-k}^\top \ldots \mathcal{M}_{j-1}^\top  \Phi,  \nu(j-k-1) \rangle. \label{formulaMarkovobservables}
\end{equation}
This formula provides the response of a general observable of the system to a general time-dependent perturbation. 
Earlier results on response of Markov chains to perturbations had been obtained under more restrictive hypotheses on the nature of the acting forcing and of the reference dynamics in the context of Ising spins \cite{Lippiello2005}. 

The special case of periodic reference dynamics is discussed in Appendix \ref{periodicMarkov}. Appendix \ref{usualformulas} shows how the classical perturbative formulas for the linear response of time-independent Markov chains derived in \cite{Lucarini2016,Santos2020,Lucarini2025} can be obtained as a special case of the results presented here. 

\section{Perturbations to Time-dependent Diffusion Processes}\label{diffusion}
We now replicate the results above for the case of non-autonomous diffusion processes. For autonomous diffusions, the response operators are  classically
expressed using correlation functions. Following the results obtained using Markov chains, we expect that for non-autonomous diffusions the response depends on the full past history of the forcing.
Let $\mathbf{x}(t)$, ${t\in\mathbb{R}}$ solve the SDE
\begin{equation}
\mathrm{d}\mathbf{x}(t) =\mathbb{E}_{\nu(n)}[\Phi]= \mathbf{b}(\mathbf{x}(t),t)\mathrm{d}t + \Sigma(\mathbf{x}(t),t)\mathrm{d}\mathbf{W}(t),\label{langevin}
\end{equation}
where  $\mathbf{x}(t)\in\mathbb{R}^{d}$, $\mathbf{b}:\mathbb{R}^{d+1}\to\mathbb{R}^d$ and $\Sigma:\mathbb{R}^{d+1}\to\mathbb{R}^{d\times d}$ are smooth functions, and $\mathrm{d}\mathbf{W}(t) \in\mathbb{R}^{d}$ is a vector whose components are the increments of $d$ independent Wiener processes. We consider the Ito convention for the noise. We assume for simplicity that Eq. \ref{langevin} describes an elliptic diffusion process, so that the diffusion matrix $D=\Sigma \Sigma^\top$ is positive definite. 
The  Kolmogorov generator is
\begin{equation}
\mathcal K_t \Phi = \mathbf{b}\cdot\nabla \Phi + \frac12 D: \nabla^2 \Phi, 
\end{equation}
where $\Phi:\mathbb{R}^d\rightarrow\mathbb{R}$ is a smooth observable, whilst $\cdot$ and $:$ indicate the scalar and Hadamard product, respectively.  The system is non-autonomous because in general both the drift $\mathbf{b}$ and the diffusion $D$ can depend explicitly on time, leading  to time-dependence in all statistical moments of the system. Specifically, the temporal evolution of density of measures (measures, henceforth, since we are considering elliptic diffusion) is described by the Fokker-Planck equation \cite{pavliotis2014book}:
\begin{equation}
\partial_t \mu(t) = \mathcal L_t \mu(t)
\end{equation}
where the generator $\mathcal{L}_t = \mathcal K_t^*$, with 
\begin{equation}
\mathcal L_t \mu = -\nabla \cdot (\mathbf{b} \mu) + \frac12 \nabla^2: D \mu.
\end{equation}
We make a hypocoercivity assumption, i.e. we assume that 
there exist $C,\lambda>0$ such that for all probability measures $\nu_1,\nu_2$,
\begin{equation}
\|P_{t,s}\nu_1  - P_{t,s}\nu_2 \|_{\mathrm{TV}} \le C e^{-\lambda(t-s)} \|\nu_1-\nu_2\|_{\mathrm{TV}},
\end{equation}
where $\|\mu \|_{\mathrm{TV}}=\sup_{\|f\|_{\infty}\leq1}\int f d\mu$ is the total variation norm and  $P_{t,s} \nu=\mathcal{T}\exp\left(\int_s^t \mathrm{d}\tau \mathcal{L}_\tau\right) \nu$, with $\mathcal{T}$ being the time-ordering operator.  Hence, the Fokker-Planck equation admits a unique equivariant measure  $\rho_0(t)$, meaning that for any initial measure $\nu$,
\begin{equation}
\lim_{s\to-\infty}  P_{t,s} \nu= \rho_0(t).\label{measurediffusion}
\end{equation}
where, roughly speaking, convergence to the equivariant measure is \textit{de facto} achieved if $t-s\gg1/\lambda$. 
The statement above - as well as Eq. \ref{covariantMarkov} obtained for Markov chains - means that if we start an ensemble of simulations in the distant past, and use the same protocol of time-dependent forcing for each member of the ensemble, after a sufficiently long time, the statistical properties of the evolved ensemble do not depend on how we initially prepared the ensemble. 

{The hypocoercivity assumption is the natural generalisation of the requirement of a non-vanishing spectral gap discussed e.g. in \cite{LucariniChekroun2023,lucarini2025generalframeworklinkingfree,zagli_SIAM:2026} for the autonomous case. Indeed, what we are requiring is that the system forgets its initial conditions exponentially fast, so that the equivariant measure is unique and physically meaningful, as it provides a statistical description of the system that does not depend on how the ensemble was prepared, as no region of phase space can remain coherent or trapped in the memory of its initial state. }

As discussed in \cite{Lucarini2017}, such a notion of well-posedness of the measure at all times is extremely relevant for coordinated climate modelling exercises like the Climate Models Intercomparison Project (CMIP; see, e.g., \url{https://wcrp-cmip.org/cmip-phases/cmip7/}) and for the recent trend of attempting the creation of so-called large ensemble simulations of state-of-the-art Earth system models \cite{Maher2021,Stevenson2023,Ye2024,Lin2025}, where the goal is to  sample more accurately the statistical properties of the changing climate over all time horizons.

For any smooth observable $\Phi$ and probability measure $\mu$ we define $\mathbb{E}_{\mu}[\Phi]=\langle \Phi,\mu\rangle=\int \mathrm{d}\mathbf{x}\mu(\mathbf{x})\Phi(\mathbf{x})$. When the system has converged to the equivariant measure, the expectation value of a measurable observable of the system changes with time as follows:
\begin{equation}
\langle \Phi \rangle_0(t)= \mathbb{E}_{\rho_0(t)}[\Phi]=\langle \Phi,\rho_0(t)\rangle. \label{observable0}
\end{equation}
We perturb the drift term as $b(\mathbf{x}(t),t)\rightarrow b^\varepsilon(\mathbf{x}(t),t)=b(\mathbf{x}(t),t)+\varepsilon g(t)\mathbf{h}(\mathbf{x}(t),t)$, so that the generators are altered as follows:
\begin{align}
\mathcal K_t^\varepsilon \Phi &= \mathcal K_t  \Phi + \varepsilon g(t) \mathcal K^{(1)}_t  \Phi, \quad  \mathcal K^{(1)}_t  \Phi= \mathbf{h}\cdot\nabla\Phi, \\ 
\mathcal L_t^\varepsilon \mu & = \mathcal L_t \mu+ \varepsilon g(t) \mathcal L^{(1)}_t  \mu, \quad \mathcal L^{(1)}_t  \mu =-\nabla ( h \mu). 
\end{align}
Another option is to consider a perturbation to the noise term $\Sigma(\mathbf{x}(t),t)\rightarrow \Sigma^\varepsilon(\mathbf{x}(t),t)=\Sigma(\mathbf{x}(t),t)+\varepsilon g(t)\Gamma(\mathbf{x}(t))$. In this case, the generators are  altered as follows
\begin{equation}
\mathcal K_t^\varepsilon \Phi = \mathcal K_t  \Phi + \varepsilon g(t) \mathcal K^{(1)}_t  \Phi+\varepsilon^2g(t)^2 K^{(2)}_t  \Phi, \label{newKolmo}
\end{equation} 
where
\begin{equation}
\mathcal K^{(1)}_t  \Phi
=1/2(\Gamma^\top \Sigma +\Sigma^\top \Gamma): \nabla^2 \Phi \quad K^{(2)}_t  \Phi = 1/2 \Gamma \Gamma^\top: \nabla^2 \Phi
\end{equation}
and
\begin{equation}
\mathcal L_t^\varepsilon \mu  = \mathcal L_t \mu+ \varepsilon g(t) \mathcal L^{(1)}_t  \mu+\varepsilon^2g(t)^2 \ \mathcal L^{(1)}_t  \mu,  \label{newPF}
\end{equation}
\begin{equation}
\mathcal L^{(1)}_t  \mu=1/2\nabla^2: (\Gamma^\top \Sigma +\Sigma^\top \Gamma) \mu \quad \mathcal L^{(2)}_t  \mu= 1/2\nabla^2:  (\Gamma \Gamma^\top  \mu),  
\end{equation}
where, for our purposes below, retaining the terms up to the first order in $\varepsilon$ in Eq. \ref{newKolmo} and Eq. \ref{newPF} is sufficient.
Let $\rho^\varepsilon (t)$ solve
\begin{equation}
\partial_t \rho^\varepsilon(t)= \mathcal L_t^\varepsilon \rho^\varepsilon(t).
\end{equation}
We assume that for sufficiently small $\varepsilon$ the hypocoercivity assumption  applies also for the $\varepsilon-$ perturbed dynamics, so that $
\lim_{s\to-\infty}  P_{\varepsilon,t,s} \nu= \rho_\varepsilon(t)$ for all probability measures $\nu$.
{We have that the unperturbed equivariant measure defined in Eq. \ref{measurediffusion} is given by:
\begin{equation}
\rho_{0}(t)=\lim_{\varepsilon\rightarrow0} \rho^\varepsilon(t), 
\end{equation}
whilst the derivative $\rho^{(1)}(t)= \left.\frac{d}{d\varepsilon}\right|_{\varepsilon=0} \rho^\varepsilon(t)$ exists} and satisfies
\begin{equation}
\partial_t\rho^{(1)}(t)= \mathcal L_t \rho^{(1)}(t)+ g(t)\mathcal {L}_t^{(1)} \rho_0(t).
\end{equation}
Using the Duhamel formula, the unique pullback-bounded
solution is
\begin{align}
\rho^{(1)}(t)= & \int_{-\infty}^\infty \mathrm{d}s \Theta(t-s) g(s) P_{t,s} \mathcal{L}_s^{(1)} \rho_0(s)\nonumber\\
=& \int_{-\infty}^\infty\mathrm{d}s\Theta(t-s) g(s)\mathcal{T}\exp\left(\int_s^t \mathrm{d}\tau \mathcal{L}_\tau\right) \mathcal{L}_s^{(1)} \rho_0(s).  \label{formulasdemeasures}
\end{align}

This is the continuous-time analog of Eq. \ref{formulaMarkovmeasures} obtained for Markov chains.
For any smooth observable $\Phi$ we have :
\begin{align}
\frac{d}{d\varepsilon}&\Big|_{\varepsilon=0} \mathbb{E}_{\rho^\varepsilon(t)}[\Phi]= \frac{d}{d\varepsilon}\Big|_{\varepsilon=0}\langle\Phi,\rho^\varepsilon(t)\rangle=\langle\Phi,\rho^{(1)}(t)\rangle\nonumber\\
&=\int_{-\infty}^\infty\mathrm{d}s\Theta(t-s) g(s)\langle\mathcal{K}_s^{(1)}\mathcal{T}\exp\left(\int_s^t \mathrm{d}\tau \mathcal{K}_\tau\right) \Phi,\rho_0(s)\rangle,\label{Greenresponsetimedep}
\end{align}
which mirrors Eq. \ref{formulaMarkovobservables} obtained for Markov chains. Note that we { can write  
\begin{equation}
\langle\Phi,\rho^{(1)}(t)\rangle=\int_{-\infty}^\infty \mathrm{d}s\Theta(t-s)\mathcal{G}_{\Phi,h}(t-s,s)g(s)\label{Greentimedep}
\end{equation}
where we have introduced the Green's function $\mathcal{G}_{\Phi,h}$ which translates the acting forcings into a time-dependent change in the expectation value of the observable. Note that the Green's function depends on both $t-s$ and $s$, as opposed to the autonomous case, where it depends only on the lag $t-s$ between the moment the forcing acts and the moment the observable is measured \cite{Santos2022,LucariniChekroun2023}. Indeed, in this case memory of the time-dependent measure is retained.} 
In the case of autonomous reference dynamics, $\mathcal{L}_t=\mathcal{L}$, $\mathcal{K}_t=\mathcal{K}$, $\rho_0(t)=\rho_0$, and the usual generalised fluctuation-dissipation results are obtained, with the Green's function losing its explicit dependence on time. The case of periodic reference dynamics is discussed in detail in Appendix \ref{periodicdiffusion}.

\section{Optimal Fingerprinting for Time-dependent Systems}\label{optimalfingerprinting}
Optimal fingerprinting aims at relating observed anomalies (with respect to a reference background) of a complex system to external forcings that perturb the system from its reference state. The task is challenging because one wants to extract information from an individual trajectory of the system. Because of the presence of natural variability within the ensemble of trajectories, the attribution of the observed anomalies to the forcings is always subject to uncertainty. The methodology described below for attributing signal anomalies to specific forcings - both natural and anthropogenic ones - has been developed in the context of climate science following the landmark contribution by Hasselmann \cite{Hasselmann1997}. In climate  science, one often assumes as reference state a steady state climate, and considers as acting perturbations both natural (e.g. volcanic and solar factors) and anthropogenic (greenhouse gases injection, aerosols, land-use change) forcings. Extremely strong periodic forcings like those associated with the daily or seasonal cycle are removed by applying filters to the output of the model runs. Such filters always have  a certain degree of subjectivity, because there is no \textit{perfect} way to remove a periodic component from an aperiodic signal.

Specifically, the optimal fingerprinting method (OFM) is a linear regression that aims at finding  the optimal weights $\beta_k$, $k=1,\ldots,K$ that allow one to realise the best reconstruction of the measured anomalies for $N$ observables $Y_j(t)$,  $j=1,\ldots,N$ using as basis $K$ fingerprints evaluated for the $N$ observables $M_{j,k}(t)$. Specifically, one aims at solving the linear regression problem:
\begin{equation}
{Y}_j(t)=\sum_{k=1}^K\mathrm{M}_{j,k}(t)\mathbf{\beta}_k+\mathbf{\nu}_j(t).\label{OFM}
\end{equation}
The  $k^{th}$ fingerprint is constructed as the average pattern of response of the system to the $k^{th}$ forcing alone. It is evaluated as an ensemble mean of many numerical simulations where the same forcing is applied but different initial conditions sampled from the reference  measure and different realisations of the natural variability are considered. Note that, by construction, OFM requires the use of models, as attribution is impossible with observational data only. Finally, the vector ${\nu}$ describes the natural variability of the  system. 
In  climate science, $Y$ is a filtered version of the observed record, which includes measurements of anomalies of climatic quantities (e.g temperature) at given locations, and the filtering is performed by taking spatial and/or temporal averages. Correspondingly, the fingerprints are targeted as such filtered quantities \cite{Allen1999,Hegerl2011,Hannart2014}. 

Depending on the specific hypotheses on the quality of the fingerprints and on the statistical properties of  the natural variability, it is possible to find different ways to solve Eq. \ref{OFM} and define best estimates and confidence level for $\beta_k$,  $k=1.\ldots,K$   \cite{Hegerl2011,Hannart2014}. Attribution of the observed anomaly to the $k^{th}$ forcing is considered successful if - most commonly - the  95\% confidence interval for the estimate of $\beta_k$ does not intersect zero. Clearly, the signal-to-noise ratio depends critically on the relative strength of the anomaly and of the natural variability. Hence, attribution may be statistically feasible only for some observables and not for others, and, clearly, by considering a filtered signal we make attribution easier by reducing the uncertainty. 

We previously showed \cite{LucariniChekroun2024}  how the OFM equations can be derived from linear response theory for statistical mechanical systems whose evolution is described by the autonomous version of Eq. \ref{langevin}, such that $\mathbf{b}(\mathbf{x}(t),t)\rightarrow \mathbf{b}(\mathbf{x}(t))$ and $\Sigma(\mathbf{x}(t),t)\rightarrow\Sigma(\mathbf{x}(t))$.  A key point that is implicitly assumed in standard OFM formulations is that all acting perturbations have to be accurately described using linear response. Another  point is that whilst OFM formulations traditionally assume that $\mathbf{\nu}$ in Eq. \ref{OFM} is the natural variability of the system in the reference state, in \cite{LucariniChekroun2024} it is explained that one should consider the natural variability of the forced state.

The results obtained in the current paper show how to derive the OFM equations in the context of non-autonomous systems, i.e. without assuming that the reference state is at steady-state. We proceed as follows. Within the setting given in Eq. \ref{langevin}, the problem above amounts to considering the quantities $\Phi_j(x^\varepsilon(t))-\langle \Phi_j,\rho_0(t)\rangle$, where $x^\varepsilon(t)$ evolves according to the perturbed equation: 
\begin{align}
\mathrm{d}\mathbf{x}^\varepsilon(t) &= \mathbf{b}(\mathbf{x}^\varepsilon(t),t)\mathrm{d}t + \sum_{k=1}^{K_1} \varepsilon_k g_k(t)h_k(\mathbf{x}^\varepsilon(t))\nonumber\\
&+\Sigma(\mathbf{x}^\varepsilon(t))\mathrm{d}\mathbf{W}(t)+\sum_{k=K_1+1}^{K} \varepsilon_k g_k(t)\Gamma_k(\mathbf{x}^\varepsilon(t))\mathrm{d}\mathbf{W}(t),
\end{align}
with some given initial condition. We consider here $K_1$ forcings acting on the drift term and $K-K_1$ forcings modifying the noise law.  The pullback measure of such a SDE can be written as $\rho^\epsilon(t)$ and from Eqs. \ref{Greenresponsetimedep}
-\ref{Greentimedep}
we have that for each observable $\Phi_j$:
\begin{equation}
\langle \Phi_j,\rho^\varepsilon(t)\rangle-\langle \Phi_j,\rho_0(t)\rangle=\sum_{k=1}^K \varepsilon_k \int_{-\infty}^\infty \mathrm{d}s\Theta(t-s)\mathcal{G}_{\Phi_j,k}(t-s,s)g_k(s)+O(\varepsilon^2),\label{Greentimedep2}
\end{equation}
where we have grouped together the $K$ acting forcings, {and where $O(\varepsilon^2)$ indicates the higher order terms of the expansion. In what follows, we neglect these terms.} Let us now write $\Phi_j(x^\varepsilon(t))=\langle \Phi_j,\rho^\varepsilon(t)\rangle+\eta_j(t)$ where $\eta_j(t)$ is a random number that describes the anomaly of the considered trajectory with respect to the ensemble average computed according to $\rho^\varepsilon(t)$. We have:
\begin{align}
\Phi_j(x^\varepsilon(t))-\langle \Phi_j,\rho_0(t)\rangle&=\delta \Phi_j(t)\nonumber\\
&=\sum_{k=1}^K \varepsilon_k \int_{-\infty}^\infty \mathrm{d}s\Theta(t-s)\mathcal{G}_{\Phi_j,k}(t-s,s)g_k(s)+\eta_j(t)\nonumber\\
&=\sum_{k=1}^K \varepsilon_k X_{j,k}(t) +\eta_j(t), \quad  j=1,\ldots N \label{fingerprints}
\end{align}
where $X_{j,k}(t)$ is the fingerprint associated with the $k^{th}$ forcing and the observable $\Phi_j$. {The procedure becomes clearer in the next section when we present a physically relevant example.} This equation has a one-to-one correspondence with Eq. \ref{OFM}, where $Y_j(t)=\delta \Phi_j(t)$, $M_{j,k}(t)=X_{j,k}(t)$ and $\nu_j(t)=\eta_j(t)$.  Indeed, 
$X_{j,k}(t)=\langle \Phi_j,\rho_k^{(1)}(t)\rangle$ where in Eq. \ref{Greentimedep2} we set $\varepsilon_q=0$ $\forall q=1,\ldots k-1, k+1,\ldots K$, thus selecting only one active forcing. 
Instead, $\eta_j(t)$ is the variability within the ensemble associated with the time-dependent measure $\rho^\varepsilon(t)$ associated with Eq. \ref{langevin}, and its covariance is time-dependent. Finally, $\varepsilon_k$ emerges naturally as the true value of the weighting factor $\beta_k$ we would infer from the data. 
Our approach shows also that it is natural to extend the OFM by considering simultaneously multiple time slices $\{t_1, t_2, \ldots, t_n\}$. Indeed, we can extend Eq. \ref{fingerprints} as follows:
\begin{equation}
\delta \Phi_j(t_m)=\sum_{k=1}^K \varepsilon_k X_{j,k}(t_m) +\eta_j(t_m), 
\label{fingerprints2}
\end{equation}
where $m=1,\ldots,n$, $j=1,\ldots N$. As a result, the statistical inference problem can be written as: 
\begin{equation}\label{finger}
\mathcal{Y}=\mathcal{M}\mathbf{\beta}+\mathcal{N}
\end{equation}
where we have 
\[
\mathcal{Y} = 
\begin{pmatrix}
Y(t_1) \\
Y(t_2) \\
\vdots \\
Y(t_n)
\end{pmatrix} \quad \mathcal{M} = 
\begin{pmatrix}
M(t_1) \\
M(t_2) \\
\vdots \\
M(t_n)
\end{pmatrix}
\quad
\mathcal{N}=
\begin{pmatrix}
\nu(t_1) \\
\nu(t_2) \\
\vdots \\
\nu(t_n)
\end{pmatrix}.
\]
The optimal value of $\beta$ is obtained as a generalised  least squares solution as:
\begin{equation}
\mathbf{\beta}_{BLUE}=\left(\mathcal{M}^\top \mathcal{S}^{-1} \mathcal{M}\right)\mathcal{M}^\top\mathcal{S}^{-1}\mathcal{Y},
\end{equation}
where $BLUE$ indicates the \textit{best linear unbiased estimator} and the relevant covariance matrix  has a block structure: 
\[
\mathcal{S} = \mathrm{Cov}(\mathcal{N},\mathcal{N})
=
\begin{pmatrix}
s_{11} & s_{12} & \cdots & s_{1n} \\
s_{21} & s_{22} & \cdots & s_{2n} \\
\vdots & \vdots & \ddots & \vdots \\
s_{n1} & s_{n2} & \cdots & s_{nn}
\end{pmatrix}
\]
where $s_{pq}\in \mathbb{R}^{N\times N}$, $s_{pq;ij}=\mathrm{Cov}(\nu_i(t_p),\nu_j(t_q))$, where Cov indicates the covariance across the ensemble. In all the formulas above - where we rely on the Gauss-Markov theorem \cite{greene2018econometric} and do not need to make assumptions of gaussianity of the natural variability - the inverse is taken in the Moore-Penrose sense \cite{Barata2012} if invertibility is lost.  Note that the resulting equations are identical to those derived for the standard case where the reference system is at steady state \cite{Hegerl2011,Hannart2014,LucariniChekroun2024}, essentially thanks to the validity of linear response in the time-dependent case considered here. 
\section{Numerical Experiments}\label{Ghil}
We now want to test the validity of the approach discussed above using as testbed a modified version of the Ghil-Sellers model \cite{Sellers,Ghil1976}, which is one of the foundational models in climate science and has played a major role in characterising the multistability of the Earth system \cite{Bodai2015,GhilLucarini2020}. We will use the model to test a) the validity of response formulas and b) the validity of the optimal fingerprinting theory developed in the previous sections.

The Ghil-Sellers energy balance model  (GSEBM) describes the processes of energy absorption, emission, and energy redistribution across the latitudes of the climate system. The model is cast as a reaction-diffusion partial differential equation describing the time evolution of the zonally-averaged surface temperature $T(x,t)$ where $x=2\phi/\pi\in[-1,1]$ is the normalised latitude $\phi$ and $t$ is time. The slightly modified version of the GSEBM we consider here can be written as follows: 
\begin{equation}\label{eq:pde}
  \begin{split} 
  \partial_tT(x,t)  &= \frac{1}{c(x)}\left(2/\pi\right)^21/\cos(\frac{\pi x}{2})\partial_x(\cos(\frac{\pi x}{2})k(x,T)\partial_xT) \\
  &+ \frac{1}{c(x)}\mu(t) Q(x)\left(1 - \alpha(x,T)\right) \\
  &- \frac{1}{c(x)}\sigma T^4\left(1 - \tilde{m}\tanh(c_3T^6)\right) +\eta_0\eta(x,t),  
  \end{split}
\end{equation} 
where standard notation of differentiation is used and with boundary and initial conditions given by  $T_x(-1,t) = T_x(1,t) = 0$ and $T(x,0) = T_0(x)$, Additionally,  $c(x)$ is the effective heat capacity of the atmosphere, land, and ocean per unit surface area at $x$. The right hand side (RHS) contains the terms describing the energy budget. The first term  represents the meridional heat transport as a diffusive law, where the diffusion coefficient $k(x,T)$ incorporates the effects of sensible and latent heat transport. 
%
The net input of solar radiation is described in the second term, which is characterised by the solar constant $\mu$, the  irradiance $Q$, and the albedo $\alpha$. The albedo decreases with temperature, as highly reflective surfaces require low temperatures. This property is associated with the ice-albedo feedback, which plays a major role for describing the multistability of the system. The longwave emission (third term on the RHS) is represented by Boltzmann's law, modified by the greenhouse effect, whose intensity is modulated by the constant $\tilde{m}$. {Large values of $\tilde{m}$ are associated with a stronger greenhouse effect.} Further details on the model are reported in \cite{Ghil1976,Bodai2015}, which we use as reference for all the tabulated functions and constants considered here. We add a stochastic forcing $\eta(x,t)$ along the lines of the Hasselmann programme \cite{Hasselmann1976,Imkeller2001,LucariniChekroun2023} in order to represent succinctly the impact of the unresolved degrees of freedom of the system {- namely, atmospheric variability and fast coupled ocean-atmosphere variability -} on the large scale, slow dynamics that is explicitly resolved in the model \ref{eq:pde}. 

We follow the same numerical implementation as in \cite{LucariniChekroun2024}. We discretise the latitude in $M=37$ steps of 5$^\circ$ and use time steps of one day.  We add at each of $j$, $j=1,\ldots,M$ site a white stochastic forcing $\eta_0\eta_j(t)$ that is uncorrelated in time and across the $M$ sites, such that, specifically, $cov(\eta_0\eta_j(t),\eta_0\eta_k(t'))=\eta_0^2\delta_{jk}\delta(t-t')$, where $cov$ indicates the covariance in time, $\delta_{ij}$ is the Kronecker delta and $\delta(t)$ is the Dirac distribution. As a result, our resulting model is an elliptic diffusion process. Following \cite{LucariniChekroun2024}, we modulate the strength of the stochastic forcing by choosing $\eta_0=0.2$, which provides  a reasonable natural high-frequency variability.  We integrate the system in time using the Euler–Maruyama scheme \cite{Kloeden2011}. 

The modification to the Ghil-Sellers model we add in this work boils down to considering a time-dependent  reference state of the system. We include an explicit time-dependent solar irradiance $\mu=\mu(t)$ in order to succinctly represent variability in the incoming solar radiation due to  solar activity and to the presence of excess aerosol in the upper atmosphere resulting from volcanic eruptions. We consider 
\begin{equation}
\mu(t)=\mu_0+\delta\mu_{SS}(t)+\delta\mu_{V}(t). \label{natural}
\end{equation}
Here, $\mu_0=1$ corresponds to  the reference solar irradiance (as used in the study \cite{LucariniChekroun2024}), whilst $\delta\mu_{SS}(t)$ aims to describe in a highly idealised manner the radiative effect of the sunspots \cite{Krivova2009,Foster2011}, so that we choose $\delta\mu_{SS}(t)=\delta_1\sin(\omega_0t)$, where $\delta_1=0.001$ and $\omega_0=2\pi/T_{SS}$, $T_{SS}=11$ $y$.

Finally, the term $\delta\mu_{V}(t)$ describes the impact of  volcanic eruptions. We parametrize $\delta\mu_{V}(t)=-\sum_{j=1}^K\alpha_j\Theta(t-t_j)\exp(-(t-t_j)/\tau)$. We select $K=13$, in order to have a volcanic eruption every $\approx 80$ $y$, and $\alpha_j\in[0.0025,0.025]$, which describes, obviously in a highly idealised fashion, events up the scale of the major Samalas (1257) and Tambora (1815) eruptions \cite{Kandlbauer2013,Stoffel2015,Wade2020}. The $t_j's$ are chosen so that the virtual volcanic eruptions occur irregularly with intervals ranging from  40 $y$ up to 120 $y$. Finally, $\tau=3y$, which is a reasonable estimate of the decay time of the impact of major volcanic eruptions on the Earth's energy budget \cite{Kandlbauer2013,Stoffel2015,Wade2020}.

{We perform multiple runs of the model, each for a duration of $1000$ $y$. We specifically choose as initial condition  the warm climate established with the present-day solar constant $\mu=1$ and $\eta_0=0$ (see Fig. 1a in \cite{Bodai2015}).  Note that we are far from the tipping point occurring for low values of $\mu=\mu_{crit}\approx0.97$ \cite{Bodai2015}, and, while we are in the region of bistability of the system, the process of noise-induced transitions from the warm to the snowball state (see \cite{Lucarini2022}) is entirely negligible given the intensity of the noise for the time-scales of interest in our investigation. {Hence, given the way the prepare our states, we consider a conditional reference equivariant measure that explores at all times  the warm states only.}}.

\subsection{Preparation of the Perturbed Ensemble}

We first perform $N_{ens}=10000$ runs, each for a duration of 1000 y, of the GSEBM in the time-dependent configuration described above in absence of global warming. Let us define as $T_{p,nat}(x,t)$ the spatial-temporal field of the $p^{th}$ ensemble member and $\langle T_{nat}(x,t) \rangle$ its ensemble average. As discussed in the paragraph following Eq. \ref{measurediffusion}, we need to ensure that we have converged to the pullback measure. To do so, we proceed as follows. We first run $N_{ens}$ simulations for the autonomous model obtained by setting $\mu(t)=1$ for 100 y, having as initial conditions the warm climate time-independent solution found in \cite{Bodai2015} for the case where stochastic forcing is switched off. This ensures that these runs populate reasonably well the invariant measure. We take the final conditions of these runs and continue the simulations using $\mu(t)=\mu_0+\delta\mu_{SS}(t)$ for a duration of $10$ $T_{SS}$= 110 $y$, which allows for the system to relax to the time-dependent pullback measure associated with the oscillating solar irradiance. We then use the final conditions of these runs as initial conditions used for the time-dependent protocol defined by Eq. \ref{natural}. We use the output obtained for the subsequent 1000 y for the statistical analyses below. The same strategy for preparing the ensembles is used also in the forced runs described below where anthropogenic forcings are included in the system.

\subsection{Computing Climate Response using Markov Chains}
Following \cite{Bodai2015,Lucarini2022}, we perform a severe coarse-graining of the model output by restricting ourselves to a two-dimensional state space $(T_{AVE},\Delta T)$. $T_{AVE}$ is the globally averaged surface temperature and is defined as  $$T_{AVE}(t)= \frac{\int_{-1}^1 \mathrm{d}x\cos(\frac{\pi}{2} x) T(x,t)}{ \int_{-1}^1 \mathrm{d}x\cos(\frac{\pi}{2} x)}.$$
Instead, $\Delta T$ is the difference between the low-latitude and high-latitude average temperature field and is defined as $$\Delta T(t)= T_E(t)-\frac{T_N(t)+T_P(t)}{2}$$ where $$T_{Q}(t)=\frac{\int_{l_{Q}}^{u_{Q}} \mathrm{d}x\cos(\frac{\pi}{2} x) T(x,t)}{ \int_{l_{Q}}^{u_{Q}} \mathrm{dx}\cos(\frac{\pi}{2} x)}$$ where $Q=\{N,S,E\}$ and $l_{S}=-1$, $l_{N}=1/3$, $l_{E}=-1/3$, $u_{S}=-1/3$, $u_{N}=1$, $u_{E}=1/3$. Additionally, we consider yearly averages instead of instantaneous fields. The (yearly-averaged) variables $(T_{AVE},\Delta T)$ are appropriate reaction coordinates because they encapsulate the key processes of the GSEBM on climatic time scales: $T_{AVE}$ controls (and is controlled by) the global energy budget, whilst $\Delta T$ controls (and is controlled by) the meridional energy transport \cite{Bodai2015,Lucarini2022}. 
\begin{figure}[htbp]
    \centering
  a)\includegraphics[width=0.6\textwidth]{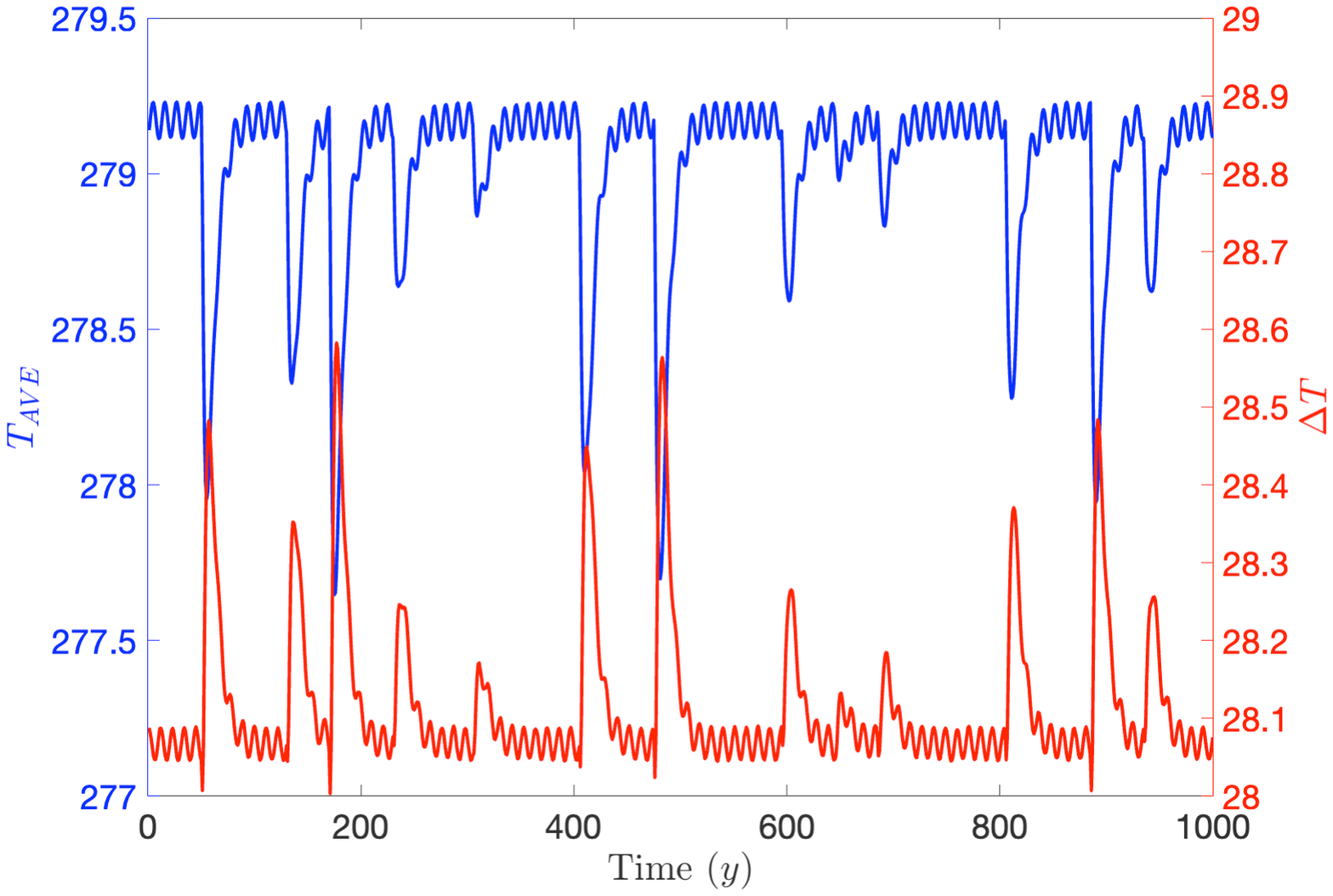}\\
  b)\includegraphics[width=0.6\textwidth]{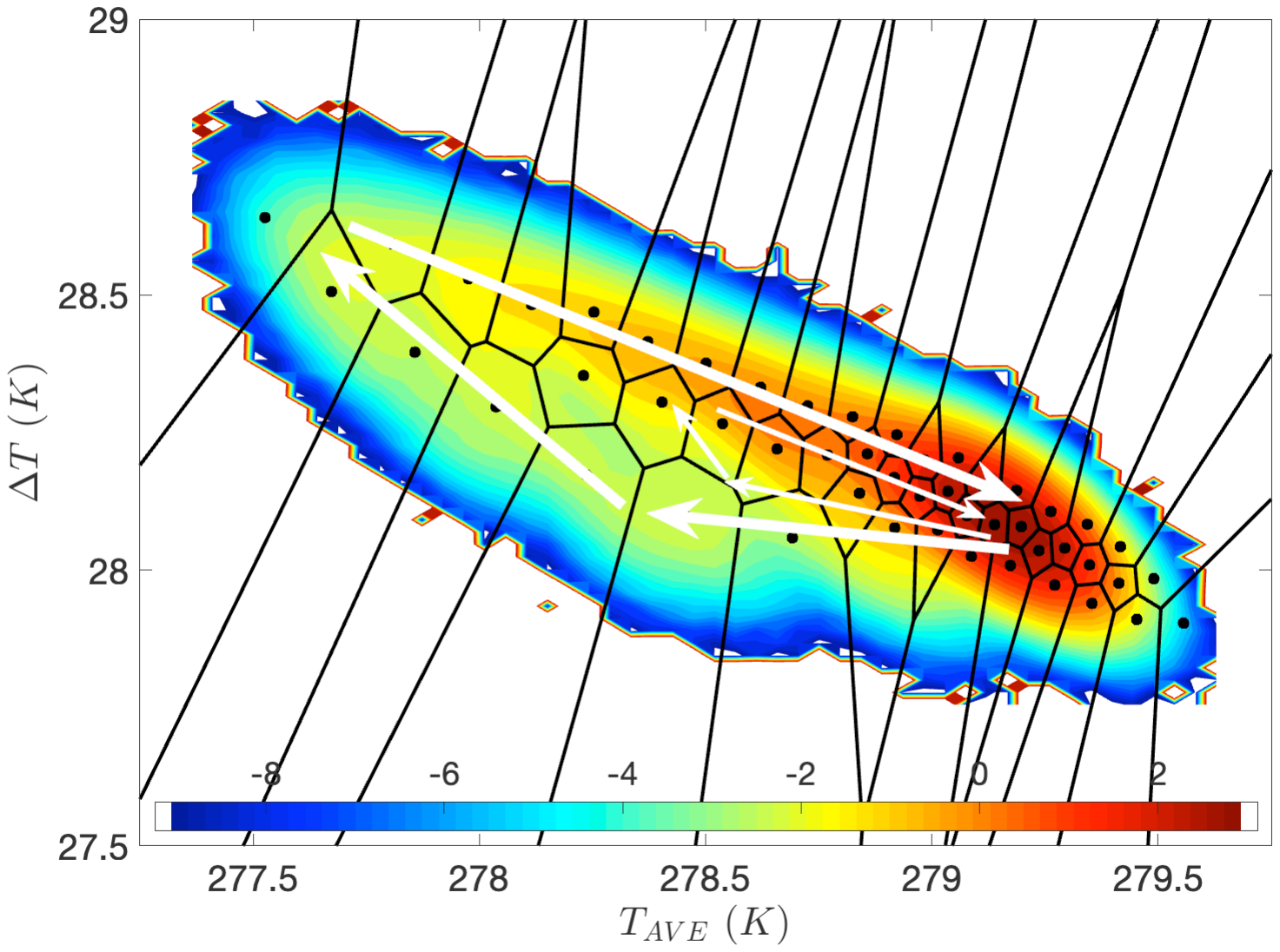}\\
  c)\includegraphics[width=0.6\textwidth]{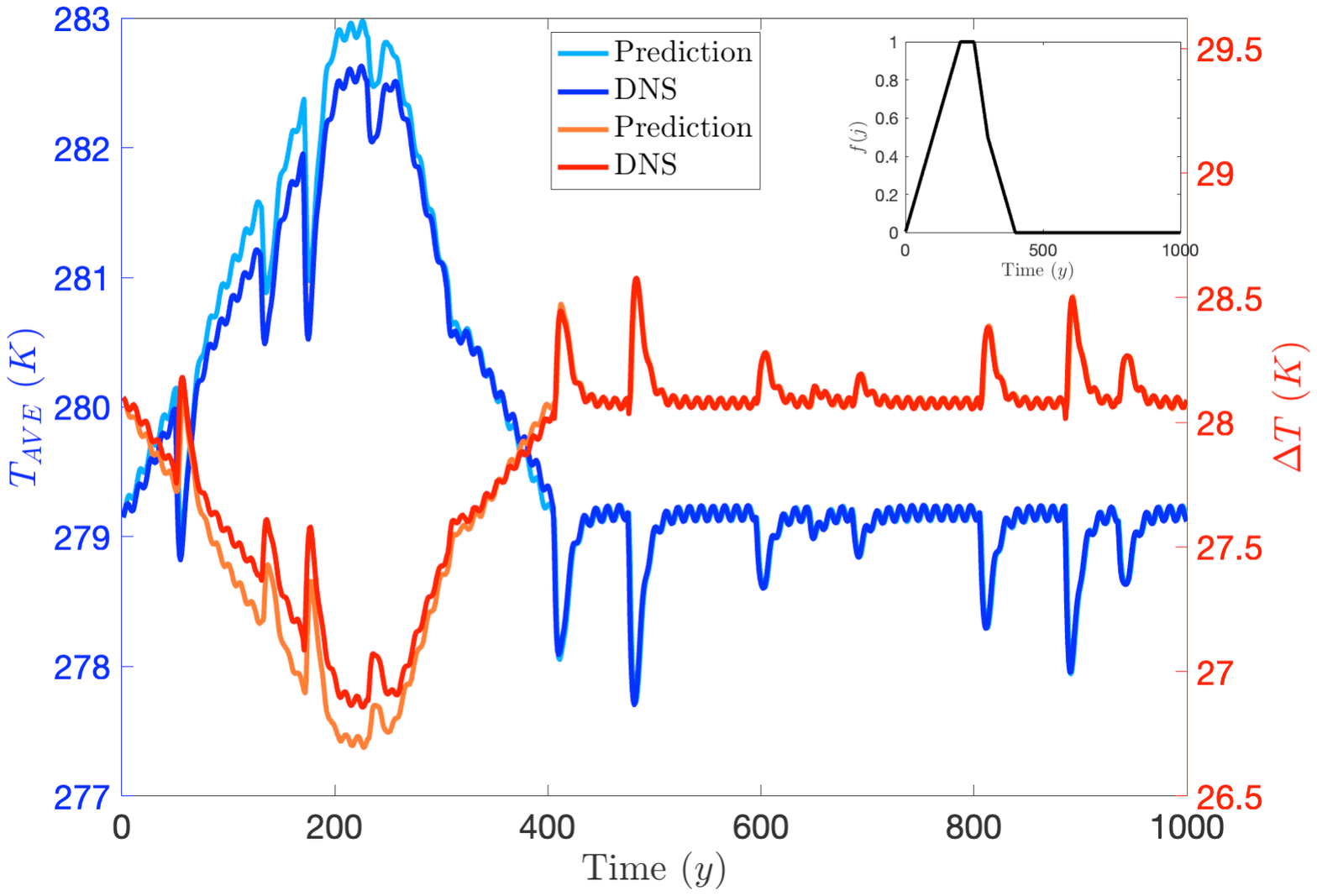}
    \caption{a) Ensemble average of simulations performed considering exclusively the sun spots cycle and volcanic eruptions as forcings to the system. b) Logarithm of the probability distribution of the yearly values of $T_{AVE}$ and $\Delta T$ for the reference time-dependent system. We have used 10000 ensemble members. The Voronoi tessellation used for constructing the reduced Markov chain is shown. The arrows provide a qualitative indication of the response of the system to the occurrence of stronger (thick arrows) or weaker (thin arrows) volcanic eruptions. c) Ensemble average of  $T_{AVE}$ (light blue) and $\Delta T$ (orange) under the anthropogenic forcing scenario where $\bar{m}$ is perturbed as according to the modulation $f(j)$ and corresponding predictions (blue and red, respectvely) obtained using the reduced Markov model constructed with the tessellation shown in b).} \label{fig:voronoi}
\end{figure}
We then perform k-means clustering \cite{Macqueen1967} on the yearly averaged values of $(T_{AVE},\Delta T)$ sampled data from all the ensemble runs and construct, using the standard Euclidean distance, a Voronoi tessellation \cite{Aurenhammer91} defining 50 states $\{B_j\}$, $j=1,\ldots,50$. 

We now construct a reduced Markov model by studying the statistics of transitions between the states $B_j$ along the lines described in \cite{Lucarini2016,Lucarini2025} Since our system is time-dependent, we need to define a different Markov model $\mathcal{M}_j$ for each of the $j=1,\ldots,1000$ years of the simulation. Using standard techniques, we first construct  the Markov model using a frequentist approach by virtue of the classical maximum likelihood estimator.  Let $n_{j;l,k}$ be number of observed transitions from state $l$ to state $k$ at time $j$ collected across the  $N_{ens}$ ensemble members
\[
n_{j;l,k} = \sum_{p=1}^{N_{ens}} \mathbf{1}_{B_l}(x^p_j) \text{ and }  \mathbf{1}_{B_k}(x^p_{j+1})
\]
where $x^p_j=(T^p_{AVE}(j),\Delta T^p(j))$ is the coarse-grained state of the system at year $j$ for the ensemble member $p$, and let $n_{j;l}=\sum_{p=1}^{N_{ens}} \mathbf{1}_{B_l}(x^p_j)$ be the total occupancy of state $l$.  The maximum likelihood estimate for $ \mathcal{M}_{j;l,k}$ is $\mathcal{M}_{j;l,k}= \frac{n_{j;l,k}}{n_{j;l}}$. 
We then use the $\mathcal{M}_{j}$'s to construct the equivariant measure $\nu(j)$ via Eq. \ref{covariantMarkov}. As a result of the latitudinal heat diffusion and of the Boltzmann feedback, the convergence to the equivariant measure is extremely rapid. 

In order to increase the robustness of our estimates, we perform a Bayesian correction for the estimate of the Markov chain $\mathcal{M}_j$. We take inspiration from \cite{Diaconis2006} and introduce a Dirichlet factor associated with creating $\sqrt{N_{ens}}=100$ pseudo-observations distributed according to the equivariant measure $\rho_0(j)$. Hence, we obtain a revised estimate of the stochastic matrix at time j,
\begin{equation}
\mathcal{M}_{j;l,k}= \frac{n_{j;l,k}+\sqrt{N_{ens}}\rho_0(j)_k}{n_{j;l}+\sqrt{N_{ens}}}.
\end{equation}
The Bayesian procedure can be iterated with no noticeable changes in the outcome. 

The time evolution of the two observables $T_{AVE}$  and $\Delta T$ computed by iterating the Markov chain above following Eq. \ref{observable0markov} is shown in Fig. \ref{fig:voronoi}a), where the effect of the aperiodic volcanic eruptions and of the periodic forcing of the sunspots is  apparent. The lines are virtually indistinguishable from those obtained by taking ensemble averages of the direct numerical simulations. Figure \ref{fig:voronoi}b) shows {\color{black}the probability distribution function of the system pooled over time and ensemble members} in the projected space with indication of the Voronoi tessellation. The arrows indicate the qualitative behaviour of the system following each volcanic eruption. We clearly see the emergence of currents {in phase space}, which are the signature of the nonequilibrium nature of the system. The GSEBM features a distinct Arctic amplification: positive anomalies in the average temperature are accompanied by negative anomalies in the temperature difference between low and high latitudes: as a result of the ice-albedo feedback, the high latitudes feature a higher sensitivity \cite{Bodai2015,Lucarini2022}. 

Next, we run a second set of $N_{ens}=10000$ simulations of the GSEBM given in Eq. \ref{eq:pde} where {we probe the greenhouse effect by applying} the perturbation $\tilde{m}=\tilde{m}_0\rightarrow \tilde{m}=\tilde{m}_0+\delta$, where $\tilde{m}_0=0.5$ and $\delta=0.0005$. We remark that this perturbation is 20 times smaller than the one considered in \cite{LucariniChekroun2024}, so we are very safely in the regime of linearity. We follow the same protocol as in the first set of experiments and, using the same states in the projected $(T_{AVE},\Delta T)$  plane  depicted in Fig. \ref{fig:voronoi}, we compute our best estimate of $\mathcal{M}^\delta_j$ and of $\nu^\delta(j)$. 

{Note that whilst both the unperturbed dynamics described by $\mathcal{M}_j$ and the perturbed dynamics described by $\mathcal{M}^\delta_j$, are time-dependent, the applied perturbation is autonomous, so that in linear approximation the difference between the two stochastic matrices is $\mathcal{M}^\delta_j-\mathcal{M}_j=\delta m_j=\delta m$. We take advantage of all yearly data of the model and we use the following estimate for the autonomous perturbation matrix: $m=\sum_{k=1}^{1000} \left(\mathcal{M}^\delta_k-\mathcal{M}_k\right)/(1000\delta)$. }

We now test the accuracy of the response theory in predicting the evolution of the expectation value of $T_{AVE}$  and $\Delta T$ when the system undergoes additional forcings. Our climate change experiments are inspired by a previous investigation \cite{LucariniChekroun2024} and involve changes in  $\tilde{m}$, according to the protocol $\tilde{m}=\tilde{m}_0\rightarrow \tilde{m}=\tilde{m}_0+\varepsilon f(t)$. {We want to describe in conceptual terms  a future climate scenario where the $CO_2$ concentration goes through a steady increase until a stabilization is achieved, followed by a progressive reduction back to preindustrial values. We choose the following time protocol:}
\begin{align}
{f}(t)=&1/200 y \times t\quad &t <200 y\nonumber \\
=&1  &200y\leq t <250 y \nonumber\\
=&1 -1/100y \times (t-250 y) \quad &250 y \leq t <300 y \nonumber\\
=&1/2 - 1/100y\times(t-300y) \quad &300 y \leq t <350 y \nonumber\\
=&0&\quad t\geq350 y,  \label{CO2}
\end{align}
where $\varepsilon=0.01=20\delta$.  All the natural forcings described in Eq. \ref{natural} are clearly also included. We realise $N_{ens}=10000$ simulation according to this protocol. We define as $T_{p,CO_2}(x,t)$ the spatial-temporal field of the $p^{th}$ ensemble member and $\langle T_{CO_2}(x,t) \rangle$ its ensemble average.

The goal now is to use our estimates of $m$, $\nu(j)$, $\mathcal{M}_j$ together with the time modulation of the forcing $\varepsilon{f}(j)$ {(where we have discretised the latter at yearly interval)} given above to predict the response of the system to the applied perturbation. The result is shown in Fig. \ref{fig:voronoi}c) for $T_{AVE}$  and $\Delta T$, where $f(j)$ is shown in the inset. We observe that despite using a severely coarse-grained - in space and time - Markov chain model, and despite using our estimate of linear response operators for predicting the impact of forcings that are 20 times as strong as those used for estimating the operators themselves, we are able to perform a rather accurate prediction, as can be seen by comparing the output of response theory with the temperature difference between low and high latitudes for the field $\langle T_{CO_2}(x,t) \rangle$ obtained from direct numerical simulations. Response theory overestimates by about 10\% the peak response, yet capturing accurately the interplay between all acting forcings, and providing good results throughout the time domain. 

\begin{figure}[htbp]
    \centering
  a)\includegraphics[width=0.55\textwidth]{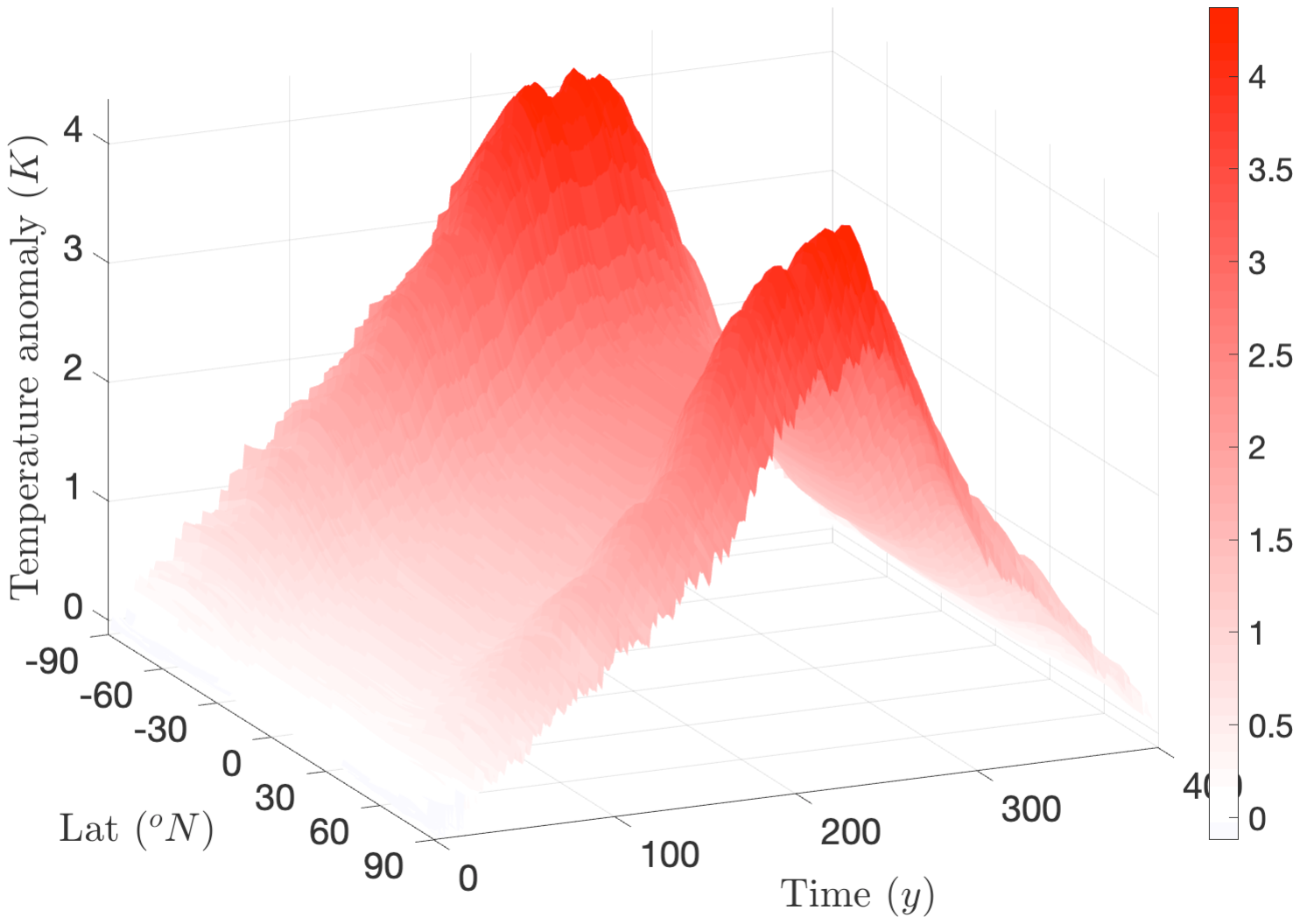}\\
  b)\includegraphics[width=0.55\textwidth]{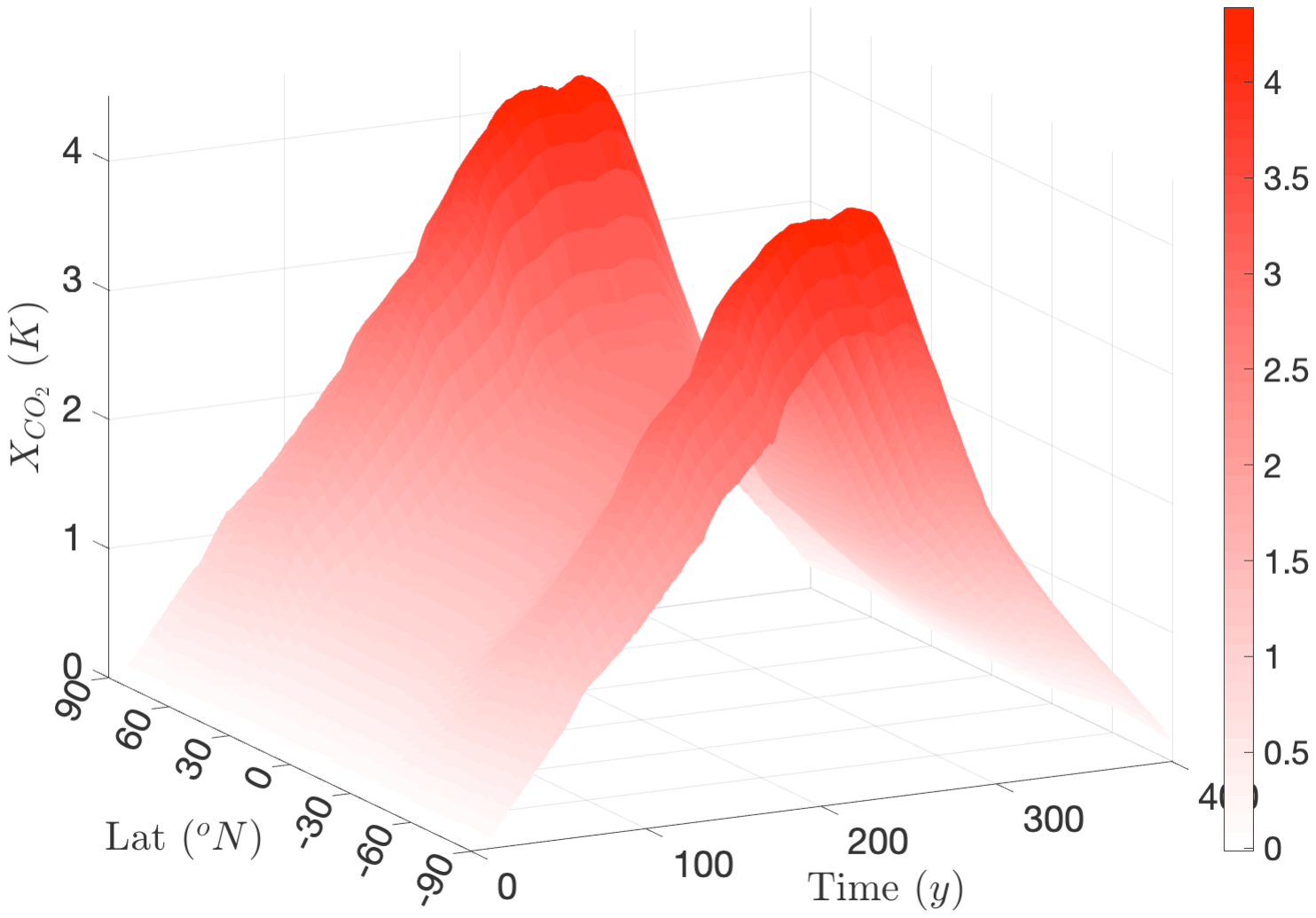}\\
  {c)}\includegraphics[width=0.55\textwidth]{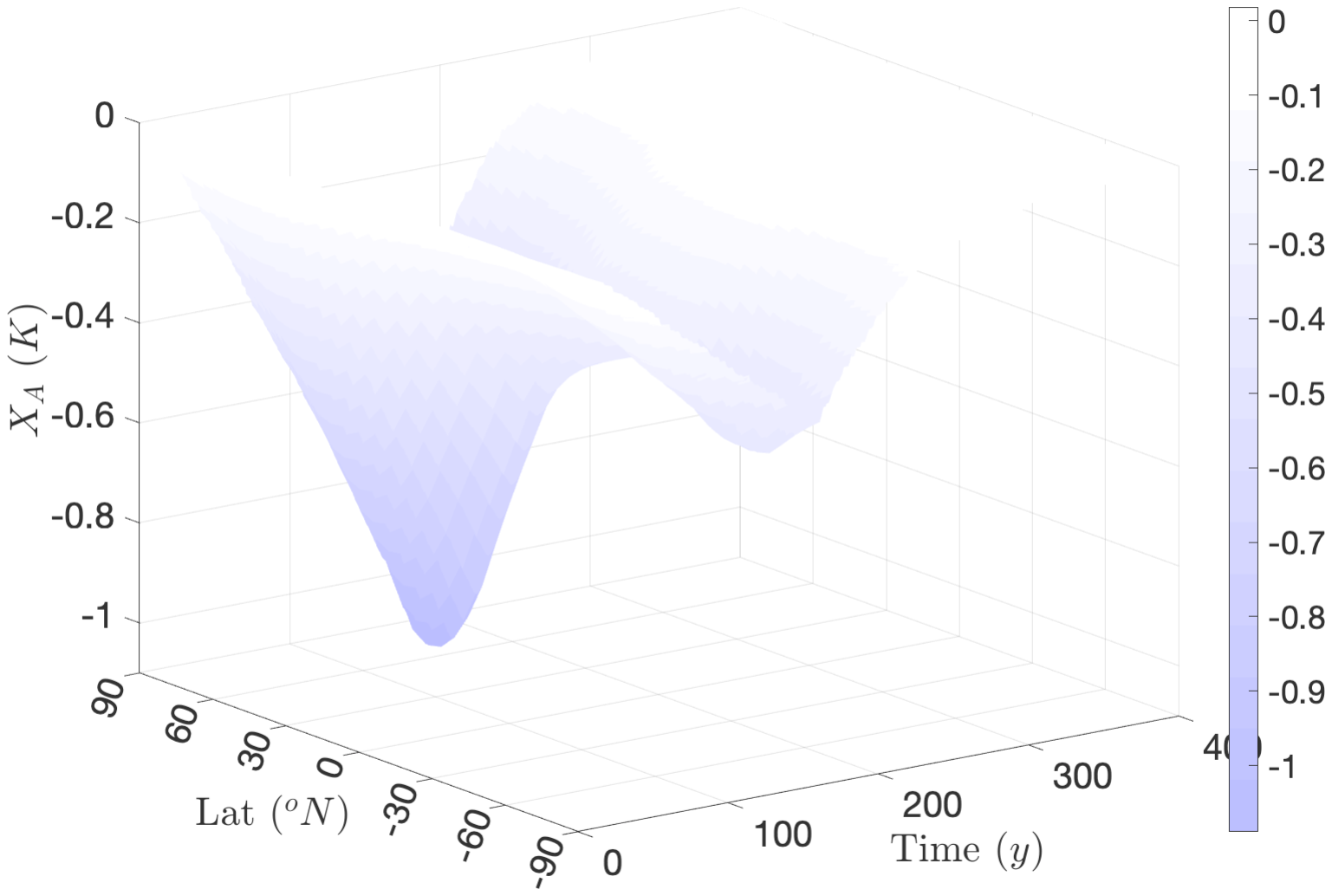}\\
    \caption{Optimal fingerprinting for detection and attribution of climate change for a nonautonomous reference state. a) Evolution of the annual anomaly of the temperature field for one ensemble member undergoing both the $CO_2$ and aerosols forcing. b) $CO_2$ forcing fingerprint $X_{CO_2}$. c) Aerosols forcing fingerprint $X_A$.} \label{fig:fingerprint}
\end{figure}

\subsection{Optimal Fingerprinting for Climate Change Detection - Nonstationary case}
As a final analysis, we wish to test the theory developed in Sect. \ref{optimalfingerprinting} using the GSEBM. In order to address a more challenging fingerprinting problem, we consider the impact of a second forcing to the system. We consider a localised reduction of the incoming radiation in the region $[25^\circ N,45^\circ N]$, mimicking the effect of anthropogenic aerosol injection in the atmosphere in the low-to-mid latitudes of the Northern Hemisphere. Such a forcing is realised by adding in the region $[25^\circ N,45^\circ N]$ a correction to the incoming solar radiation. Specifically, the forcing we apply to Eq. \ref{eq:pde} is of the form $\varepsilon_1\gamma(t)/c(x)\mathbf{1}_{[x_1,x_2]}(x)Q(x)/\left(1 - \alpha(x,T)\right)$ where $\varepsilon_1=-0.012$, $[x_1,x_2]=[5/18,1/2]$ refers to the spatial domain where the forcing is applied, and
\begin{align}
\gamma(t)=&1/150 y \times t\quad &t <150 y\nonumber \\
=&\exp(-(t-150y)/50y \quad &t\geq 150 \label{AE}
\end{align}
 is the time modulation. The forcing increases linearly in 150 years, and then fades away exponentially with decay time of 50 years. This describes conceptually a prototypical IPCC technology transition scenario, where aerosol emissions first peak and then decay as more advanced technological options become available. We generate $N_{ens}=10000$ simulations where the natural forcings described by $\mu(t)$ plus the aerosol forcing is active. We then generate $N_{ens}=10000$ simulations where we consider the natural forcings described by $\mu(t)$ plus the anthropogenic forcings associated with $CO_2$ and aerosols concentration changes. We define as $T_{p,A}(x,t)$ the field of the $p^{th}$ ensemble member and $\langle T_{A}(x,t) \rangle$ its ensemble average.

By definition - see Sect. \ref{optimalfingerprinting} - $X_{CO_2}(x,t)=\langle T_{CO_2}(x,t) \rangle-\langle T_{nat}(x,t) \rangle$ is the $CO_2$ fingerprint and $X_{A}(x,t)=\langle T_{A}(x,t) \rangle-\langle T_{nat}(x,t) \rangle$ is the aerosols fingerprint, where we take ensemble averages.

We now perform $N_{ens}=10000$ simulations where we include the natural forcing, the $CO_2$ forcing, and the aerosols forcing. Let us define as $T_{p,all}(x,t)$ the spatial-temporal field of the $p^{th}$ ensemble member of such a set of simulations. The signal for which we want to apply the OFM is $Y_p(x,t)=T_{p,all}(x,t)-\langle T_{nat}(x,t)\rangle$, i.e. the anomaly observed with respect to the (time-dependent) ensemble average of the system in absence of any anthropogenic forcing. In Fig. \ref{fig:fingerprint} we portray an example of temperature anomaly field for an example member (Panel a), the $CO_2$ fingerprint (Panel b), and the aerosols fingerprint (Panel c). The $CO_2$ fingerprint is quantitatively dominant, so that there is no obviously recognisable signature of the aerosols forcing in the temperature anomaly signal. 

We next perform for each ensemble member $p$ and at each time between $t=1$ $y$ and $t=400$ $y$ the linear regression - see Eq. \ref{finger} - of the signal $Y_p(x,t)$ with respect to $X_{CO_2}(x,t)$ and $X_{A}(x,t)$, thus obtaining a matrix of values $(\beta_{CO_2}(p,t),\beta_{A}(p,t))$, $t=1,\ldots,400$ $y$ and $p=1,\ldots,N_{ens}$. We area-weight the temperature at latitude $\pi x/2$ through the factor $\cos(\pi x/2)$ in order to take into account the metric of the Earth surface. This allows us to estimate empirically ensemble averages, variances, and covariances of the $(\beta_{CO_2}(p,t),\beta_{A}(p,t))$ dataset. 

\begin{figure}[htbp]
    \centering
  a)\includegraphics[width=0.66\textwidth]{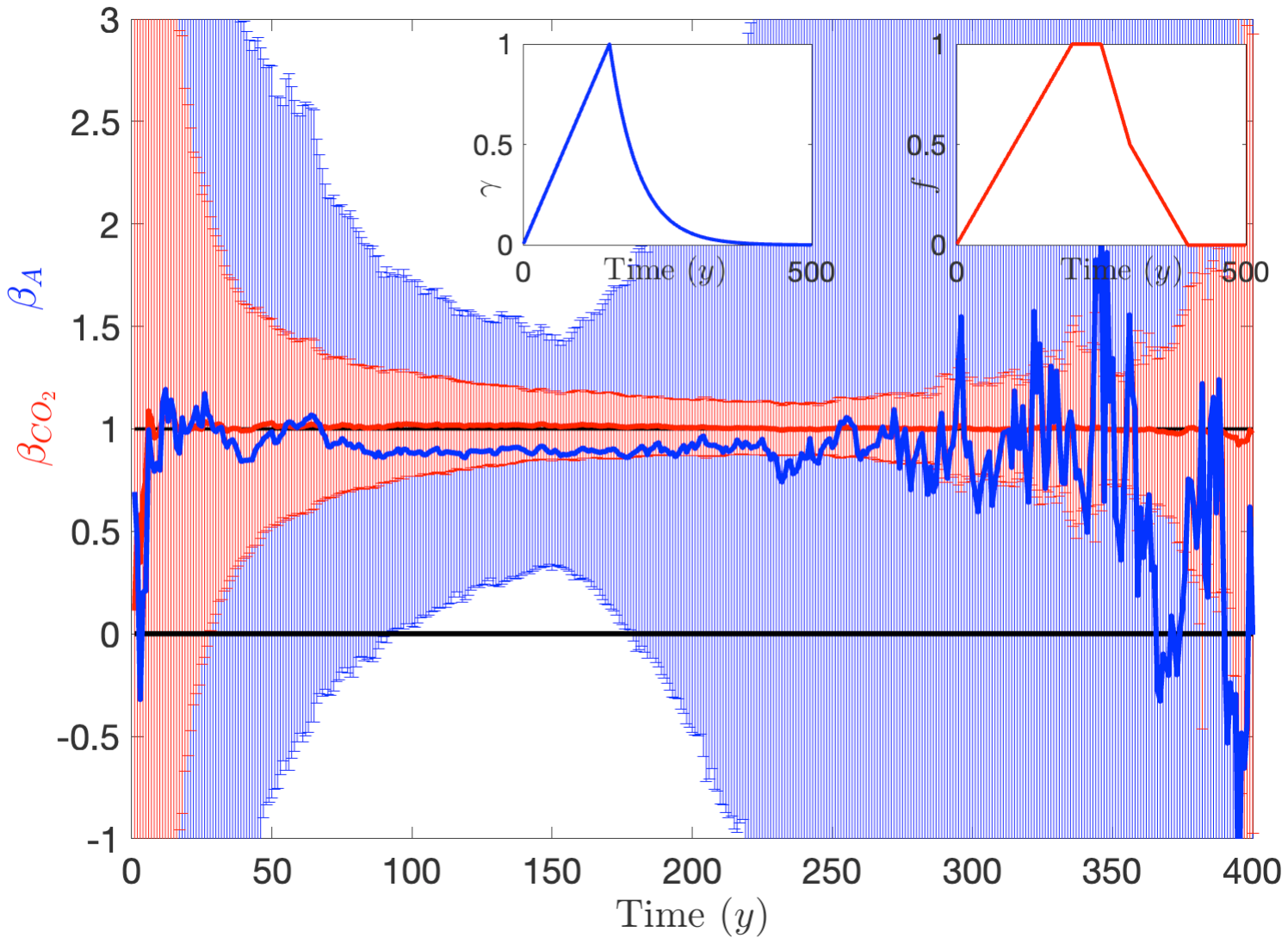}\\
  b)\includegraphics[trim={2cm 7cm 2cm 8cm},clip,width=0.33\textwidth]{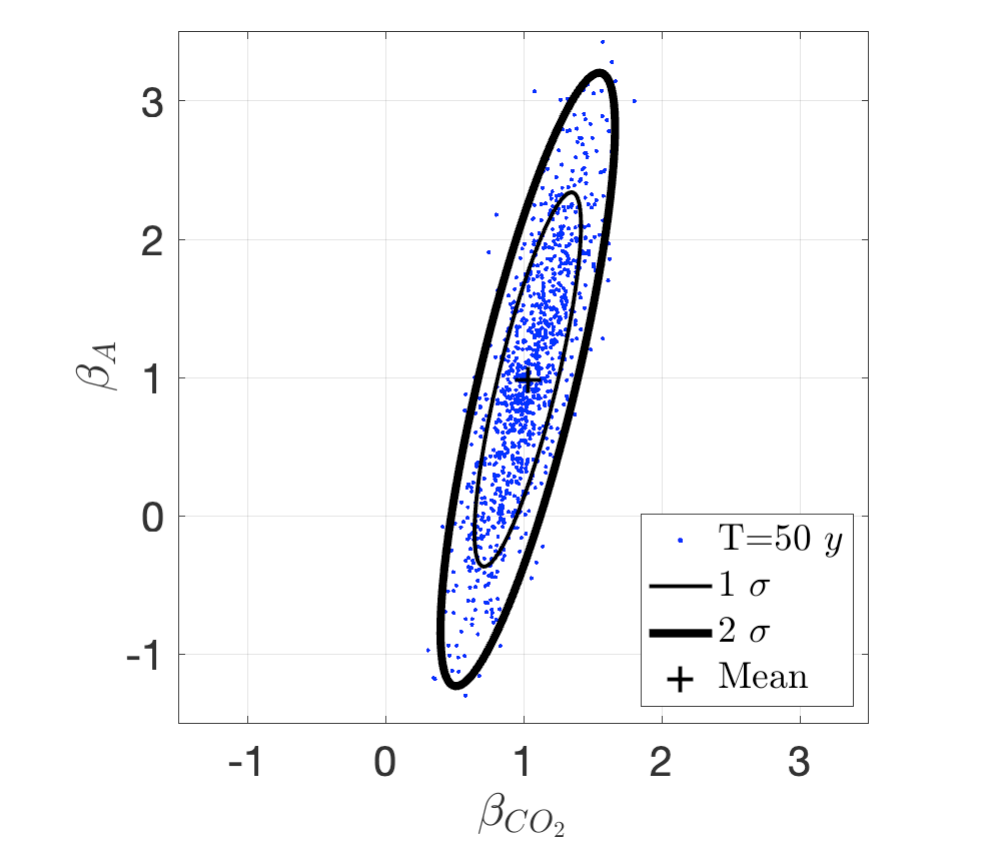}  c)\includegraphics[trim={2cm 7cm 2cm 8cm},clip,width=0.33\textwidth]{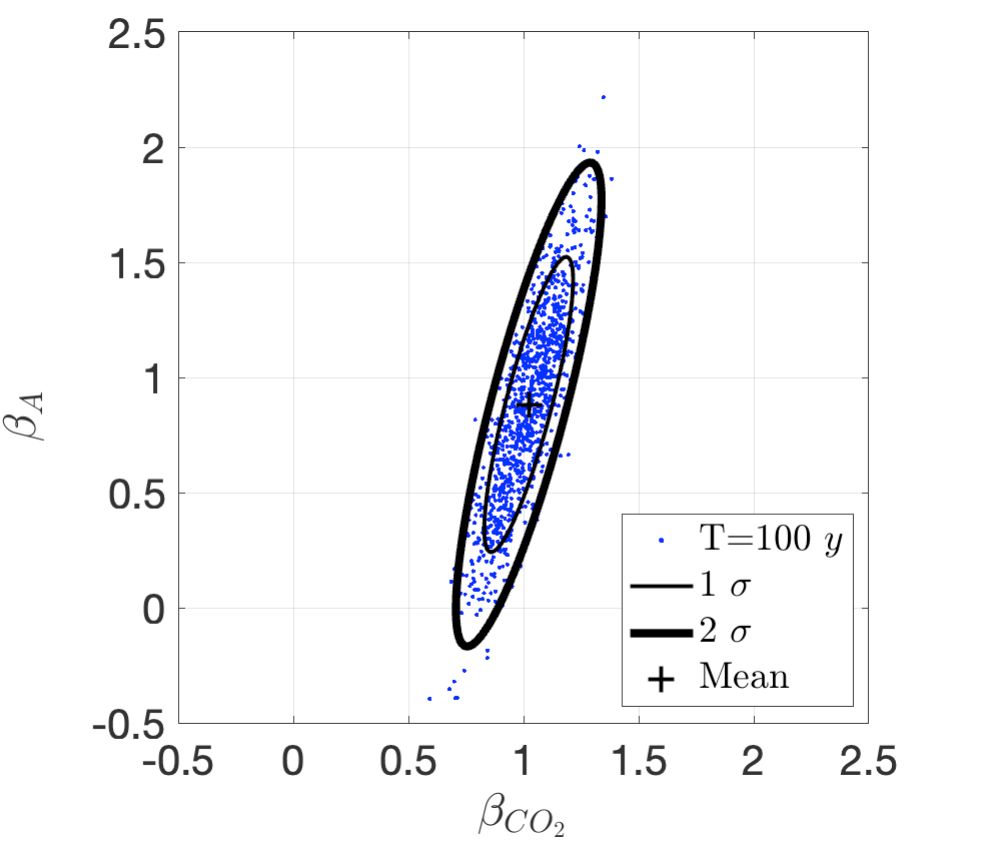}\\
   d)\includegraphics[trim={2cm 5.5cm 2cm 7cm},clip,width=0.33\textwidth]{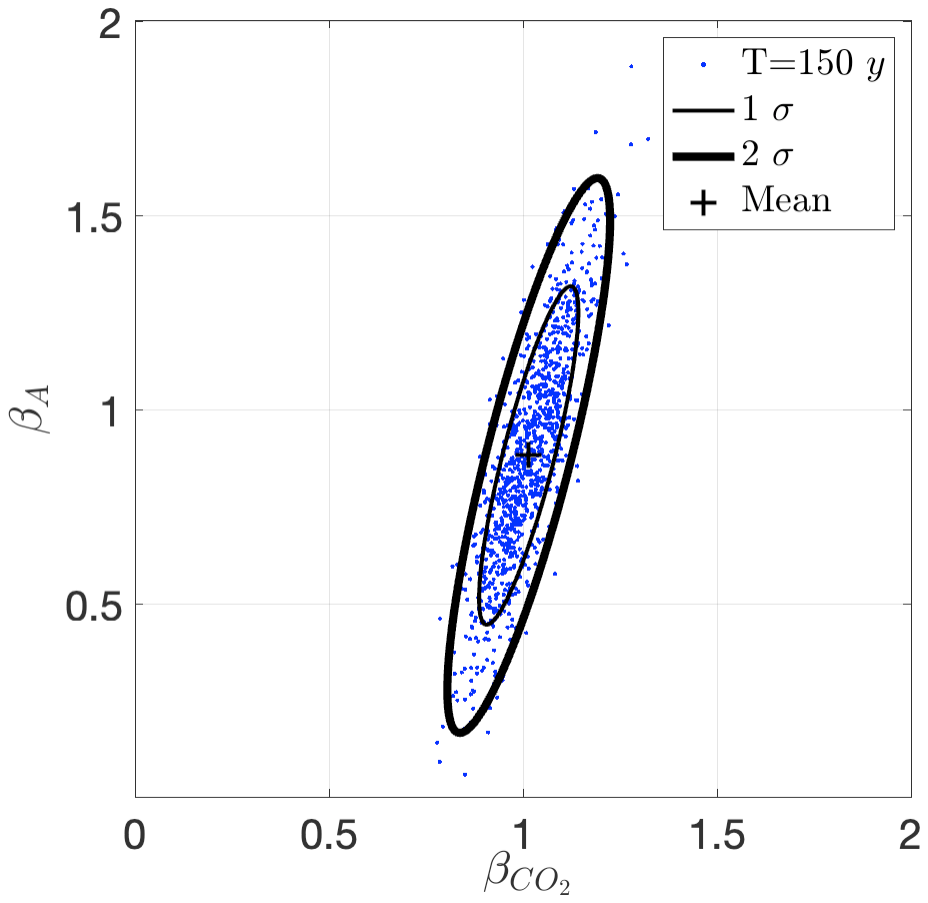}  e)\includegraphics[trim={2cm 6cm 2cm 8cm},clip,width=0.33\textwidth]{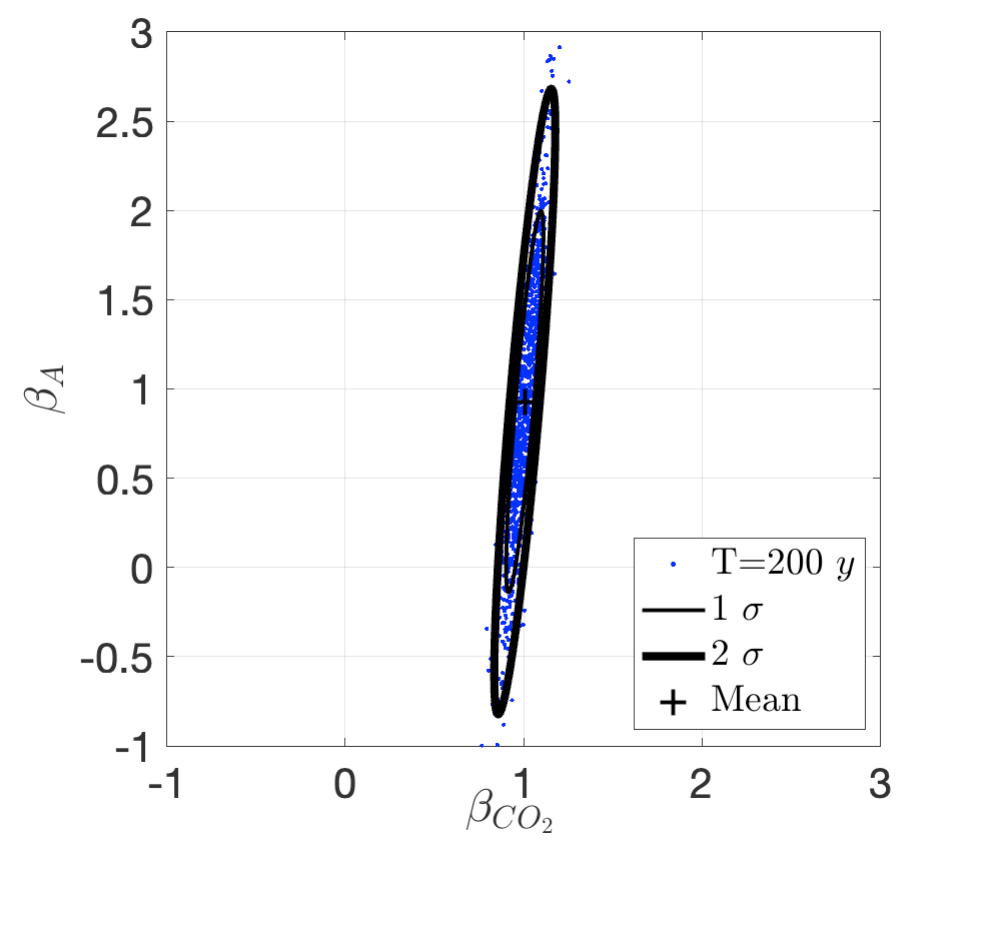}
    \caption{Optimal fingerprinting for detection and attribution of climate change for a nonautonomous reference state. a) Average value $\pm$ 2 standard deviations computed across the ensemble simulations of the weighting factor the $ CO_2$ ($\beta_{CO_2}$) and aerosols ($\beta_A$) fingerprints. The corresponding temporal modulations of the forcing $f(j)$ (for CO$_2$) and $\gamma(j)$ (for aeroslos) are shown in the insets. b) Scatter plot of the $\beta$ factors with indication of the 1 and 2 $\sigma$'s confidence region at the time horizon T=50 $y$. c) Same as b), for  T=100 $y$. d) Same as b), for  T=150 $y$. e): Same as b), for  T=200 $y$.} \label{fig:OFM}
\end{figure}

The key results are shown in Fig. \ref{fig:OFM}. We drop the $p-$ dependence on the $\beta$ quantities as we discuss ensemble statistics. The confidence interval for $\beta_{CO_2}(t)$ does not intersect zero in the time range  $t\in [\approx 30y, \approx370y]$. This means that in this range we have positive attribution of the observed climate change (with respect to the time-dependent natural variability) to the $CO_2$ forcing. The confidence interval is rather stringent between in the time range $t\in [\approx 100y, \approx330y]$, indicating that high confidence in the attribution is possible also far away from the peak $CO_2$ forcing, which is realised for $t\in [200y, 250y]$. For $t>400y$ the attribution exercise makes no sense because no anthropogenic $CO_2$ forcing is active. Note that at almost all times the ensemble mean of $\beta_{CO_2}(t)$, $t\in [\approx 30y, \approx370y]$ is very close to unity, indicating that the optimal fingerprinting method gives the ideal outcome. 

In the case of the aerosol forcing, the attribution exercise is, unsurprisingly, more challenging, for the fundamental reason that we are trying to extract the fingerprint of a much weaker forcing, as already noted in Fig. \ref{fig:fingerprint}a.  Attribution of the observed anomaly signal to aerosols forcing is statistically significant only for a couple of decades around $t=150$ $y$, which corresponds to the peak of the forcing. For $t>200$, the aerosols forcing is evanescent, hence the confidence interval of the optimal fingerprinting becomes very large.

There is a strong positive correlation between the estimates of $\beta_{CO_2}(t)$ and of $\beta_{A}(t))$ - see panels b)-e) in Fig. \ref{fig:OFM} - for the basic reason that the two forcings have opposite effects on mean average surface temperature. Nonetheless, the aerosol forcing is asymmetric between two hemispheres, and this is the key aspect that makes it possible to identify it when performing the optimal fingerprinting because we consider the full surface field in our detection procedure. 

\section{Conclusions}\label{conclusions}
In this paper we have tried to bring together three exciting notions - namely a) linear response theory \cite{Ruelle1998GeneralLinearresponseformula,Ruelle2009,HairerMajda2010,Santos2022,LucariniChekroun2023}, b) equivariant measures and snapshot attractors \cite{Crauel1997,CHEKROUN20111685,Bodai2013,GhilLucarini2020,Drotos2015,Herein2016,Tel2020,Bodai2020,Janosi2021}, and c) optimal fingerprinting for detection and attribution of climate change \cite{Hasselmann1997,Allen1999,Hegerl2011}. This complements and extends previous attempts in this direction. In \cite{Lucarini2017} we proposed to use response theory to practically construct the pullback measure of system that are weakly perturbed from a non-equilibrium steady state. In \cite{LucariniChekroun2024} we showed that linear response theory provides solid dynamical foundations for the state-of-the art OFM.

Here we have embraced time-dependence and abandoned the notion of steady state reference state. We have generalised response formulas for studying how a time-dependent system responds to extra forcings. We have considered systems whose asymptotic state is described by a pullback measure which is explicitly time dependent. Ensembles initialised in the distant past converge exponentially fast to such a measure, which defines the reference state. We have performed the calculations in two separate classes of system, namely finite state Markov chains and diffusion processes. In the latter case, we have been able to treat the case of perturbation to either the drift or the noise term. We have highlighted a clear correspondence between the formulas obtained in the two settings. 

The results obtained in the case of diffusion processes - and specifically Eqs. \ref{Greenresponsetimedep}-\ref{Greentimedep} as well all the derivation pertaining to the OFM - can be formally and heuristically extended to time-dependent chaotic systems, whose dynamics is supported at each instant on a fractal snapshot attractor \cite{Romeiras1990,Bodai2013}, even if we deem necessary to perform a  careful dedicated analysis.   

 As a result of the explicit time-dependence of the reference system, response operators do not have the form of the classical convolution product between the time modulation of the forcing and a Green's function that does not depend explicitly on time and that determines the delayed impact of the extra forcing on the system. In this case, whilst causality is preserved, the Green's function depends explicitly on time.  Our results reduce to what are now classical results of nonequilibrium statistical mechanics in the case that the reference system is at steady-state and has a unique invariant measure. In the limit that the time modulation of the reference dynamics is infinitely slow, our results provide the adiabatic limit. 
We have also been able to describe perturbatively how the presence of a time modulation in the reference measure impacts the linear response.  Finally, in the special but relevant case that the pullback measure is periodic, every Green's function inherits such a property, so that it is possible to reconstruct the full response operators of the systems from a limited set of probe experiments.

An outcome of the theory developed in this paper is the possibility of defining optimal fingerprinting also for a time-dependent system, i.e. to perform a statistical analysis of the anomaly signal obtained from an individual trajectory aimed at linking causally such anomaly to an acting cause, namely an extra forcing added to the system. Optimal fingerprinting has become a powerful and extremely valuable tool for performing detection and attribution of climate change, by assuming a hypothetical steady state climate as preindustrial  reference state. Thanks to our approach, we can e.g. include natural forcings as part of the reference dynamics and treat anthropogenic forcings alone as extra perturbations. Time-dependent components of the reference dynamics can be associated with aperiodic perturbation (e.g. volcanic eruptions), periodic components (e.g. daily and seasonal cycle, astronomical and astrophysical modulations), or, in the case one treats a specific climatic subcomponent like the atmosphere (the ocean), slowly changing forcings associated with the coupling with the ocean (the ice). Since the climate is a complex system featuring natural variability over a vast range of time scales, our approach allows for a better physically grounded optimal fingerprinting method, able to deal with the effect of multiple sources of natural climate variability.

The more flexible methodology proposed here allows for performing fingerprinting and linking causally signal and forcings also for systems that not even approximately near a steady state.  We can treat the case of systems whose reference state is described by non-autonomous components that cannot be treated perturbatively - as in the case of strong periodic forcing or of sequences of singular perturbations. Hence, we foresee applications in areas like neurosciences, biology, polymers, finance, and  quantitative social sciences. {Additionally, our results allow one to simultaneously apply the fingerprinting method  to multiple time slices, thus allowing for a global - in time - identification of the forcings.} 

In order to give more grounding to our results, we have applied the methodology developed in this paper to the investigation of the response of a variant of the Ghil-Sellers energy balance model where we introduce fairly nontrivial natural forcings associated with the sun spot cycle and volcanic eruptions. As a result of such forcings, there is no reference steady state. Following \cite{LucariniChekroun2024}, we introduce additional perturbations that are reminiscent of anthropogenic forcings, namely $CO_2$ increase and aerosols injection in the midlatitudes. We show that the response formulas we have developed in this paper are able to predict accurately the response of the system to $CO_2$ forcing even we study the problem using a severely coarse-grained approach, namely constructing a discrete Markov chain along the lines of Markov state modelling. Additionally, we have tested successfully the validity of the optimal fingerprinting method developed in this paper in identifying the two acting forcings.

In previous investigations, using spectral methods and linking response theory with Koopmanism, we had linked the divergence of the linear response to the occurrence of so-called bifurcation-induced tipping phenomena, showing that thus formalism allows us to extend classical results to high-dimensional complex systems and provides solid foundations to the theory of critical slowing down \cite{LucariniChekroun2023,LucariniChekroun2024,Lucarini2025,lucarini2025generalframeworklinkingfree}. In future works we will explore where the results presented in this paper allow to provide a probabilistic and statistical mechanical view on critical phenomena of rate-induced tipping \cite{Ashwin2012,Panahi2023} and phase-induced tipping \cite{Alkhayuon2021}, which are receiving a great deal of attention across multiple scientific communities. {Indeed, our generalised Green's functions framework allows one to test the sensitivity of the response  to perturbations initiated at different phases or times of the reference evolution.}
The results presented here might also be useful for identifying accurate early warning signals. 

There is a growing trend of casting response theory as an optimization problem whereby one selects the optimal (according to some cost function) perturbation for achieving a  goal, which is typically a desired change in the value of an observable of interest \cite{Castro2011,Galatolo_2017,MacKay_2018,Antown2018,Antown2022,Gutierrez2025,Dambrosia2026,froyland2026}. This angle has the merit of reversing response theory and casting it as a bottom-up rather than the usual top-down methodology. It is intriguing to consider how to such a hybridization of response and control theory can be extended to the case of time-dependent reference dynamics. 

The results presented in this paper are mostly formal and would require a more rigorous mathematical treatment. This has so far been achieved in the case of time-discrete dynamics in a companion paper \cite{Galatolo2026}, {whilst a second companion paper is devoted to studying the convergence of the Ulam-like approximations for the equivariant measure and for the linear response in the limit of increasingly fine resolution of the phase space \cite{galatolo2026b}. In this latter paper a numerical example shows convincingly that, even if the hypocoercivity assumption proposed in this paper might seem overly restrictive, linear response theory can apply even if the time modulation of the reference state is very strong, leading to fundamental changes in the properties of the equivariant measure.}. A complete proof of linear response theory for continuous time-dependent stochastic process is still elusive, but hopefully within reach.  Better understanding the key mathematical conditions that are necessary for deriving response formulas is also key to anticipating the realm of their practical applicability.  

Another limitation of the analysis presented here we have not discussed how to optimally estimate the various ingredients of the OFM from models and data, and we have set ourself in the not-so-realistic perfect model scenario. These aspects are indeed extremely important for ensuring practical applicability of our results \cite{Hannart2014}. Extending our work to address these issues is another direction we will pursue in the future. 

Finally, we plan to apply the methodology presented in this paper to more complex models of climate and to other complex systems - e.g. ecosystems, neural systems, networks, finance - where the time modulation of the reference state can be very strong.

\section*{Acknowledgments}
VL wishes to acknowledge useful interactions with L. Giorgini, J. Moroney, M. Santos Gutierrez,  and N. Zagli. Special thanks to M. Branicki, who got the author interested in studying the response to time-dependent systems and offered some key insights. VL acknowledges the partial support provided by the Horizon Europe Projects Past2Future (Grant No. 101184070) and ClimTIP (Grant No. 100018693), by the ARIA SCOP-PR01-P003 - Advancing Tipping Point Early Warning AdvanTip project, by the European Space Agency Project PREDICT (Contract 4000146344/24/I-LR), and  by the NNSFC  International Collaboration Fund for Creative Research Teams (Grant No. W2541005).

\section*{Data Availability}
The data used for generating the figures included in this paper and the MATLAB\cite{MATLAB2024} codes used for running the numerical model and processing its output can be accessed here \cite{lucarini_data_2026}.

\appendix
\section{Periodic Reference Dynamics - Markov Chains}\label{periodicMarkov}
Assume that we have a periodic reference dynamics of period $T$ defined by $\mathcal{M}_{n+T}=\mathcal{M}_n$ $\forall n$, which leads to a periodic equivariant measure $\nu(n+T)=\nu(n)$ $\forall n$. Sometimes periodic systems are not classified as time-dependent because they are time independent
when viewed stroboscopically with period $T$. 
We show below that the response satisfies a forced linear recurrence with periodic coefficients. Define the monodromy operator
\[
P_n^T = \mathcal{M}_{n+T-1}\cdots \mathcal{M}_{n}.
\]
where clearly $P_n^T=P_{n+qT}^T$ $\forall q\in\mathbb{Z}$. The response can be regrouped using a Floquet-like representation by splitting the pullback sum into blocks of size 
T:
\begin{align}
\nu^{(1)}(n) &=\sum_{p=0}^{\infty} \sum_{j=1}^{T} \mathcal{M}_{n-1}\ldots \mathcal{M}_{n-pT-j+1}
\mathcal{M}_{n-pT-j} m_{n-pT-j-1} \nu(n-pT-j-1)\nonumber \\
&=\sum_{p=0}^{\infty} P_{n-1}^p   \sum_{j=1}^{T} \mathcal{M}_{n-1}\ldots \mathcal{M}_{n-j}
m_{n-pT-j-1} \nu(n-j-1).\label{periodic1}
\end{align}
Finally, we can obtain the response formulas for the observable $\Phi$ as follows:
\begin{align}
&\frac{d}{d\varepsilon}\Big|_{\varepsilon=0} \mathbb{E}_{\nu^\varepsilon(n)}[\Phi] = \langle \Phi, \nu^{(1)}(n)\rangle \\
&= \sum_{p=0}^{\infty}  \sum_{j=1}^{T}\langle m_{n-pT-j-1}^T \mathcal{M}_{n-k}^T \ldots  \mathcal{M}_{n-1}^T \left(P_{n-1}^p\right)^T   \Phi,  \nu(n-j-1) \rangle.\label{periodic2}
\end{align}
If the forcing is of the form $m_{j}=mf(j)$, it is possible to learn how the system responds to perturbations of arbitrary time modulation by considering a limited set of targeted simulations.
Let us assume that $f(j)=\delta_{j,-1}$. We obtain
\begin{align}
\langle \Phi, \nu^{(1)}(n)\rangle =    \langle \mathcal{M}^T_{n-1}\ldots \mathcal{M}^T_{n-j^*}\left(P_{n-1}^{p^*}\right)^T \Phi, \nu(-1)\rangle,
\end{align}
where $n=p^*T+j^*$. Let us now choose $f(j)=\delta_{j,-l}$, with $l=1,\ldots T$, we have that 
\begin{align}
\langle \Phi, \nu^{(1)}(n)\rangle =    \langle \mathcal{M}^T_{n-1}\ldots \mathcal{M}^T_{n-j^*}\left(P_{n-1}^{p^*}\right)^T \Phi, \nu(-l)\rangle,
\end{align}
where $n=p^*T+j^*-l+1$. As a result, by performing $T$ experiments with $l=1,\ldots,T$, and taking advantage of the periodicity of the underlying dynamics, we are able to reconstruct all the terms needed for predicting the response of the system to an arbitrary time modulation of the forcing. Each of the $T$ terms constructed as above provides the contribution of the corresponding phase of the oscillation of the measure supported on the pullback attractor. Hence, even if also in this case the response cannot be written as the convolution of a Green's function with the time modulation of the forcing, one can reconstruct the operator by tailoring probing perturbations.

This makes clear why, if the underlying dynamics is aperiodic or even quasiperiodic (so that we can approximately treat it as periodic with period $T\rightarrow\infty$), it is hard to reconstruct the time-dependent response of the system to arbitrary perturbations from a limited set of experiments.

\section{Recovering the Perturbation Formulas for Autonomous Markov Chains}\label{usualformulas}
We have not required that the equivariant measure is  \textit{close}  to a suitably defined invariant measure. Here we show that if instead this is the case, it is possible to recover, when suitable limits are  taken, previously derived response formulas. 

Let us assume that $\mathcal{M}_j=\mathcal{M}+\zeta q_j$, where $\{q_j\}$, $n\in\mathbb{Z}$ are signed matrices whose columns have vanishing sum and $\zeta\in\mathbb{R}$ such that $\mathcal{M}_j$ is a stochastic matrix $\forall j\in\mathbb{Z}$. Let us define  $\nu_\zeta(j)$ the limit sequence of measures defining the pullback attractor of this system, see Eq. \ref{pullback}. From \cite{Lucarini2025}, we have that if a perturbative expansion can be obtained, we have that $\nu_\zeta(j)=\nu_0+\sum_{l=1}^\infty\zeta^l\delta^{(l)}\nu(j)$, where $\delta^{(l)}\nu(j)$ defines the $l^{th}$ perturbative term and where $\nu_{\zeta=0}(j)=\nu_{inv}$, where $\mathcal{M}\nu_{inv}=\nu_{inv}$. 

We consider the perturbation $\mathcal{M}_j\rightarrow\mathcal{M}_j+\varepsilon m f(j)$ and expand Eq. \ref{formulaMarkovobservables} in powers of $\zeta$: 
  \begin{align}
&\frac{\mathrm{d}}{\mathrm{d}\varepsilon}\Big|_{\varepsilon=0} \mathbb{E}_{\nu_\zeta^\varepsilon(j)}[\Phi] =\sum_{k=-\infty}^{\infty}\mathcal{G}_{m,\Phi}(k)  f(j-k-1)\\
&+ \zeta {\mathrm{d}\zeta}\Big|_{\zeta=0} \Big(\sum_{k=-\infty}^{\infty}\Theta(k) \langle m_{j-k-1}^T (\mathcal{M}_{j-k}+\zeta q_{j-k})^T \ldots \nonumber\\
&(\mathcal{M}_{j-1}+\zeta q_{j-1})^T  \Phi,  \nu_\zeta(j-k-1) \rangle \Big) +O(\zeta^2). \label{formulaMarkovobservables2}
\end{align}
The first term on the right hand side delivers  the classical convolution formula where $\mathcal{G}_{m,\Phi}(k)=\Theta(k) \langle m^T (\mathcal{M}^T)^k \Phi,  \nu_{inv}\rangle$ is the causal Green's function associated with the observable $\Phi$, the perturbation transition matrix $m$, and the reference stationary measure $\nu_{inv}$ already presented in \cite{Lucarini2025}. The higher-order terms (in powers of $\zeta$) describes the correction to the linear response to the $m$ perturbations due to the non-stationarity of the reference measure.

It is helpful to recollect that if $\zeta=0$, choosing $f(j)=\delta_{j,-1}$, where $\delta_{p,q}$ is the Kronecker's delta, we have $\mathrm{d}/\mathrm{d}\varepsilon|_{\varepsilon=0} \mathbb{E}_{\nu_{\zeta=0}^\varepsilon(j)}[\Phi] =\mathcal{G}_{m,\Phi}(j)$, so that with a single experiment we learn how the system reacts to perturbations with arbitrary time modulations. 
\section{Periodic Reference Dynamics - Diffusion Processes}\label{periodicdiffusion}
If the reference dynamics is periodic with period $T$, so that $\mathcal{L}_{t+T}=\mathcal{L}_t$ and $\mathcal{K}_{t+T}=\mathcal{K}_t$ $\forall t$, we introduce the monodromy operator $\Pi_t=P_{t,t-T}=\mathcal{T}\exp\left(\int_{t-T}^t \mathrm{d}\tau \mathcal{L}_\tau\right)$. Equation \ref{formulasdemeasures} reads as: 
\begin{align}
\rho^{(1)}(t)&= \sum_{k=0}^\infty \int_{t-T}^t  \mathrm{d}s g(s-kT) \Pi^k_t \mathcal{T}\exp\left(\int_{s}^t \mathrm{d}\tau \mathcal{L}_\tau\right) \mathcal{L}^{(1)} \rho_0(s)\label{formulasdemeasuresperiodic}
\end{align}
whilst from Eq .\ref{Greenresponsetimedep} one obtains:
\begin{equation}
\langle\Phi,\rho^{(1)}(t)\rangle =\sum_{k=0}^\infty \int_{t-T}^t \mathrm{d}s g(s-kT)\langle\mathcal{K}^{(1)}\mathcal{T}\exp\left(\int_s^t \mathrm{d}\tau \mathcal{K}_\tau\right)  \left(\Pi^k_t\right)^*\Phi,\rho_0(s)\rangle\label{formulasdeobservablesperiodic},
\end{equation}
which correspond to Eqs. \ref{periodic1} and \ref{periodic2} obtained for Markov chains, respectively. 

Note that also in this case
$$
\langle\Phi,\rho^{(1)}(t)\rangle=\int_{-\infty}^\infty \mathrm{d}s\Theta(t-s)\mathcal{G}_{\Phi,h}(t-s,s)g(s)
$$
but here, thanks to the periodicity of the reference state, $\mathcal{G}_{\Phi,h}(t-s,s)=\mathcal{G}_{\Phi,h}(t-s,s+T)$ $\forall s$.

Let us choose $g(t)=\delta(t-t_0)$, $0\leq t_0<T$. We have that $t=t_0+nT+\tau$, where $0\leq\tau<T$. The response at time $t$ is:
\begin{align}
&\langle\Phi,\rho^{(1)}(t)\rangle=\langle \mathcal{K}_{t_0}^{(1)}\mathcal{T}\exp\left(\int_{t_0}^{t_0+\tau} \mathrm{d}\tau' \mathcal{K}_{\tau'}\right)  \left(\Pi^n_t\right)^*\Phi,\rho_0(t_0)\rangle\label{formulasdeobservablesperiodic2}
\end{align}
If we repeat the experiment with $0\leq t_0<T$, and we observe the response at time $t$, we are able to reconstruct the full response operator. As discussed above in the case of Markov chains, if, instead, the background dynamics is time independent, the response operator coincides with the response of the system to a forcing modulated by $g(t)=\delta(t)$. This procedure follows closely the construction of the extended phase space discussed in \cite{Branicki2021}. 

\def\cprime{$'$} \def\cprime{$'$} \def\cprime{$'$} \def\cprime{$'$}
  \def\cprime{$'$} \def\cprime{$'$} \def\cprime{$'$} \def\cprime{$'$}
  \def\Rom#1{\uppercase\expandafter{\romannumeral #1}}\def\u#1{{\accent"15
  #1}}\def\Rom#1{\uppercase\expandafter{\romannumeral #1}}\def\u#1{{\accent"15
  #1}}\def\cprime{$'$} \def\cprime{$'$} \def\cprime{$'$} \def\cprime{$'$}
  \def\cprime{$'$} \def\cprime{$'$} \def\cprime{$'$}
  \def\polhk#1{\setbox0=\hbox{#1}{\ooalign{\hidewidth
  \lower1.5ex\hbox{`}\hidewidth\crcr\unhbox0}}} \def\cprime{$'$}
  \def\cprime{$'$} \def\cprime{$'$}

\end{document}